\documentclass{aa}  
\usepackage{txfonts}
\usepackage{amsmath}	
\usepackage{xcolor}
\usepackage{orcidlink}
\definecolor{verde}{rgb}{0.,0.56,0.}
\definecolor{lightblue}{rgb}{0.1,0.6,0.93}
\definecolor{mblue}{rgb}{0, 0.5, 0.815}
\definecolor{RedStrong}{rgb}{0.85, 0.1, 0.1} 
\usepackage{comment}
\definecolor{purple}{rgb}{0.5,0,0.87}

\usepackage{natbib}
\usepackage{hyperref}
\usepackage{pifont}
\usepackage{graphicx,dblfloatfix}
\usepackage{adjustbox}
\usepackage{booktabs}
\usepackage{dblfloatfix}
\usepackage{float}

\renewcommand*{\bibfont}{\small}
\hypersetup{colorlinks, linkcolor=cyan, citecolor=blue, urlcolor=purple} 

\begin{document} 
    \title{How large can galaxies be?}
    \subtitle{Ultra-deep imaging of IC~1101, the most extended known galaxy}

\author{
Carlos Marrero-de la Rosa\inst{1,2} \thanks{Corresponding author: carlos.marrero.astro@gmail.com} \orcidlink{0009-0005-1728-8076} \and 
Ignacio Trujillo \inst{1,2} \orcidlink{0000-0001-8647-2874} \and 
Ignacio Ruiz Cejudo \inst{1,2} \orcidlink{0009-0003-6502-7714} \and
Mireia Montes \inst{6}  \orcidlink{0000-0001-7847-0393} \and
Fernando Buitrago \inst{3,4,5} \orcidlink{0000-0002-2861-9812} \and 
Giulia Golini \inst{7} \orcidlink{0009-0001-2377-272X} \and
Sergio Guerra Arencibia \inst{1,2} \orcidlink{0009-0001-7407-2491} \and
Andrés Asensio Ramos \inst{1,2} \orcidlink{0000-0002-1248-0553}\and
Raúl Infante-Sainz \inst{8} \orcidlink{0000-0002-6220-7133} \and
Adriana de Lorenzo-Cáceres\inst{1,2} \orcidlink{0000-0002-9744-3486} \and
Jairo Méndez-Abreu\inst{1,2} \orcidlink{0000-0002-8766-2597} \and
Samane Raji \inst{3} \orcidlink{0000-0001-9000-5507} \and
Javier Román \inst{9} \orcidlink{0000-0002-3849-3467} \and
Manuel Sánchez-Benavente \inst{1,2} \orcidlink{0000-0003-2693-279X} \and
Zahra Sharbaf \inst{6} \orcidlink{0009-0004-5054-5946} 
}

\institute{
Instituto de Astrof\'isica de Canarias, c/ V\'ia L\'actea s/n, E-38205 - La Laguna, Tenerife, Spain
\and
Departamento de Astrof\'isica, Universidad de La Laguna, E-38205 - La Laguna, Tenerife, Spain
\and
Departamento de Física Teórica, Atómica y Óptica, Universidad de Valladolid, 47011, Valladolid, Spain
\and
Laboratory for Disruptive Interdisciplinary Science (LaDIS), Universidad de Valladolid, 47011 Valladolid, Spain
\and 
Instituto de Astrofísica e Ciéncia do Espaço, Universidad de Lisboa, OAL, Tapada de Ajuda, PT1349-018 Lisbon, Portugal
\and
Institute of Space Sciences (ICE, CSIC), Campus UAB, Carrer de Can Magrans, s/n, 08193 Barcelona, Spain
\and
School of Physics, University of New South Wales (UNSW Sydney), Sydney, NSW 2052, Australia
\and
Centro de Estudios de Física del Cosmos de Aragón (CEFCA), Plaza San Juan 1, E-44001 Teruel, Spain
\and
Departamento de Física, Universidad de Córdoba, Campus de Rabanales, Edificio Albert Einstein,
E-14071 Córdoba, Spain
}

\date{\today}

\abstract{
The maximum physical extent that galaxies can reach is poorly understood. In this regard, IC 1101, one of the most extended and massive galaxies known, provides a valuable opportunity to constrain the upper limit of galaxy sizes at the present epoch. Previous deep imaging of the system confirmed its enormous extension, but did not indicate whether it has an edge. We explore this issue using the deepest images ever taken of this galaxy; ultra-deep \textit{g}- and \textit{r}-band imaging from the INT/WFC, reaching $\mu \sim 30$ mag arcsec$^{-2}$ (3$\sigma$ in an area equivalent to $10 \times 10 \ \mathrm{arcsec}^2$). We model and subtract the scattered light from both stars and the galaxy itself using an extended PSF characterization and a hybrid wavelet-based deconvolution. Using a combination of surface brightness, colour, and stellar mass density profiles oriented at different position angles, we find that the main body of IC 1101 extends to $R_{\mathrm{edge}} \approx 260$ kpc along the semi-major axis (assuming the redshift of Abell 2029, $z = 0.077$), enclosing $\sim3.4 \times 10^{12} M_\odot$ in stars. This $R_{\mathrm{edge}}$ is among the largest edge radii measured for any galaxy to date, placing IC 1101 at the extreme upper end of the mass–size relation. In addition, we report a large number of asymmetrical, very low surface brightness features around the galaxy that are spatially consistent with the large-scale disturbances observed in the intracluster medium through X-ray studies of the Abell 2029 cluster, in which IC 1101 is embedded. With a confirmed diameter of around 520 kpc, IC 1101 stands as the largest galaxy known to date; yet, its outskirts show clear signatures of ongoing mass assembly, indicating that its spatial extent is still growing.
}
\keywords{galaxies: evolution -- galaxies: formation -- galaxies: interactions -- galaxies:
photometry -- galaxies: individual: IC~1101 -- methods: data analysis -- methods: numerical --
methods: observational}

\maketitle
%

\section{Introduction}\label{sec:introduction}

The sizes of known galaxies vary enormously, ranging from small, ultra-faint compact satellites with extensions of approximately 15 pc \citep{cerny2026} to the most massive elliptical galaxies, which can be several hundred  kpc in diameter. The size of a galaxy is closely linked to the stellar mass it contains \citep[e.g.,][]{shen2003, bernardi2014}. This relation is approximately well described by a power law of the following form: $R\propto M^{1/3}$ \citep[e.g.,][]{hall2012,trujillo2020}. Growth is also linked to the fusion of different astronomical objects and to the formation of new stars from gas  \citep[i.e. bottom-up structure formation, e.g.,][]{naab2009,oser2010,rodriguez-gomez2016}. Consequently, galaxy size is also a function of cosmic age \citep[e.g.,][]{vanderwel2014, buitrago2024}.
Therefore, a natural question arises: what are the largest galaxies that the Universe has produced to date? Given the relation between stellar mass and galaxy size, we should focus on the most massive systems to determine how extended the largest galaxies can be nowadays.

The most massive galaxies today are the brightest cluster galaxies (BCGs). They represent the most extreme outcomes of hierarchical galaxy formation. These systems are thought to grow predominantly through the repeated accretion of smaller galaxies and groups, a process often described as galaxy cannibalism, which operates efficiently in the dense environments of galaxy clusters \citep[e.g.,][]{ostriker1975,hausman1978,delucia2007,trujillo2011,pillepich2018,Edwards2020,remus2022}. Both observational studies and cosmological simulations indicate that a significant fraction of the stellar mass of BCGs is assembled at late times through dissipationless mergers, leading to the gradual build-up of extended stellar envelopes \citep[e.g.,][]{bernardi2007,hopkins2009,mendez-abreu2012,rodriguez-gomez2016}. As a result, BCGs can reach spatial extents well beyond those of typical massive galaxies, developing diffuse halos that blend smoothly into the intracluster stellar component \citep{delucia2007,rudick2011,laporte2013,contini2014}. In this context, BCGs are not only products of their host clusters’ dynamical histories, but also signposts of the most advanced stages of galaxy evolution.

While BCGs provide a natural laboratory to investigate the extent of galaxy outskirts, their environment makes it quite difficult to precisely estimate their edges. Their outskirts, shaped by the cumulative effect of minor mergers and the tidal stripping of infalling satellites, preserve the fossil records of past interactions and the assembly history of the host halo \citep[e.g.,][]{Cooper2015,spavone2017}. Also, the stellar halos of BCGs can overlap with the intracluster light \citep[ICL; see reviews by][]{Contini2021review,montes2022}, further complicating the distinction between galaxy- and cluster-scale stellar components. For these reasons, we aim to select a BCG that is already identified as a massive and extended system with a relatively coherent morphology, indicative of a largely virialised state. Focusing on such systems helps to minimise ambiguities associated with ongoing mergers or complex environments, as their structure is expected to be more dynamically settled and therefore easier to characterise in a consistent way. Among known BCGs with these properties, IC~1101 stands out as an exceptionally massive and extended galaxy. It is located at the centre of the massive galaxy cluster Abell~2029 \citep[$M_{200} = 8.5\times10^{14}\,M_\odot$,][]{Walker2012,Sohn2019} at $z = 0.077$ \citep{Sohn2019}. Early deep R-band photometry by \citet{uson1991} revealed its unusually large projected extent, tracing stellar emission out to $\sim$607~kpc, placing IC~1101 among the most extended stellar systems ever observed. They derived an integrated luminosity of $10^{12}\ \,L_\odot$, making IC~1101 one of the brightest known BCGs \citep{paranjape2012,lan2016}. The galaxy also hosts one of the largest galactic depleted cores observed in a BCG \citep[$r_c \sim 4.2$~kpc;][]{dullo2017}.

In order to establish the size limits of such extraordinary system, in this work, we present a detailed analysis of IC~1101 based on ultra-deep $g$- and $r$-band imaging obtained with the Wide Field Camera (WFC) at the Isaac Newton Telescope (INT). The primary objective of this study is to characterise the full spatial extent of IC~1101 and assess its stellar mass distribution down to unprecedented (i.e. $\sim$30 mag arcsec$^{-2}$) surface brightness levels. Special attention is given to the detection of asymmetric features and sharp structural transitions in the galaxy outskirts, which may be associated with group-scale accretion events or residual signatures of past mergers \citep[e.g.,][]{Deason2021}.

This paper is organised as follows. In Section~\ref{sec:data}, we describe the observations and data reduction of the INT/WFC imaging. Section~\ref{sec:methods} presents the methodology adopted throughout this work, including the construction of the extended Point Spread Function (PSF), the subtraction of foreground stars, the application of wavelet-based deconvolution techniques. In Section~\ref{sec:radial_structure}, we analyse the surface brightness, colour, and stellar mass surface density profile, as well as the radial structure of IC~1101, identifying and characterising the different structural transitions detected in the profiles. Section~\ref{sec:size_comparison} introduces several observational size tracers to quantify the spatial extent of the galaxy and presents the corresponding stellar mass estimates. The implications of our results are discussed in Section~\ref{sec:discussion}, and our main conclusions are summarised in Section~\ref{sec: conclusions}.

Throughout this paper, we assume a flat $\Lambda$CDM cosmology with $\Omega_{\mathrm{m}} = 0.3$, $\Omega_{\Lambda} = 0.7$, and $H_0 = 70\,\mathrm{km\,s^{-1}\,Mpc^{-1}}$. At the redshift of Abell~2029 ($z = 0.077$; \citealt{Sohn2019}), this corresponds to a physical scale of 1~arcsec $\approx 1.46$~kpc. All magnitudes are given in the AB system.

\section{Data} \label{sec:data}

IC~1101 was observed using the WFC mounted on the 2.5 m INT, located at the Roque de los Muchachos Observatory. The observations were conducted in the Sloan $g$- and $r$-bands, as part of program 078-INT10/22A (PI: Fernando Buitrago), carried out between 27–30 May 2022. The WFC is a mosaic camera composed of four CCDs, providing a wide field of view of approximately $34' \times 34'$ with a pixel scale of $0.33''$/pixel. The $g$-band data consist of 79 individual exposures of 180 seconds each, while the $r$-band data comprise 53 frames of 300 seconds, with total exposure times of 3.95 hours and 4.41 hours, respectively. The final co-added image covers a sky region of approximately $55' \times 66'$, with a pixel scale of $0.33''$/pixel (corresponding to $\sim 0.50$ kpc/pixel at the distance of IC~1101). The resulting images reach limiting surface brightness levels of 30.5 mag arcsec$^{-2}$ in the $g$-band and 30 mag arcsec$^{-2}$ in the $r$-band, measured at a significance level of $3\sigma$ within an area equivalent to $10 \times 10 \ \mathrm{arcsec}^2$. The average seeing of the final co-added images is $\sim1.7''$ in the $g$-band and $\sim1.4''$ in the $r$-band.

\begin{figure*}[t]
    \centering
    \includegraphics[width =\linewidth]{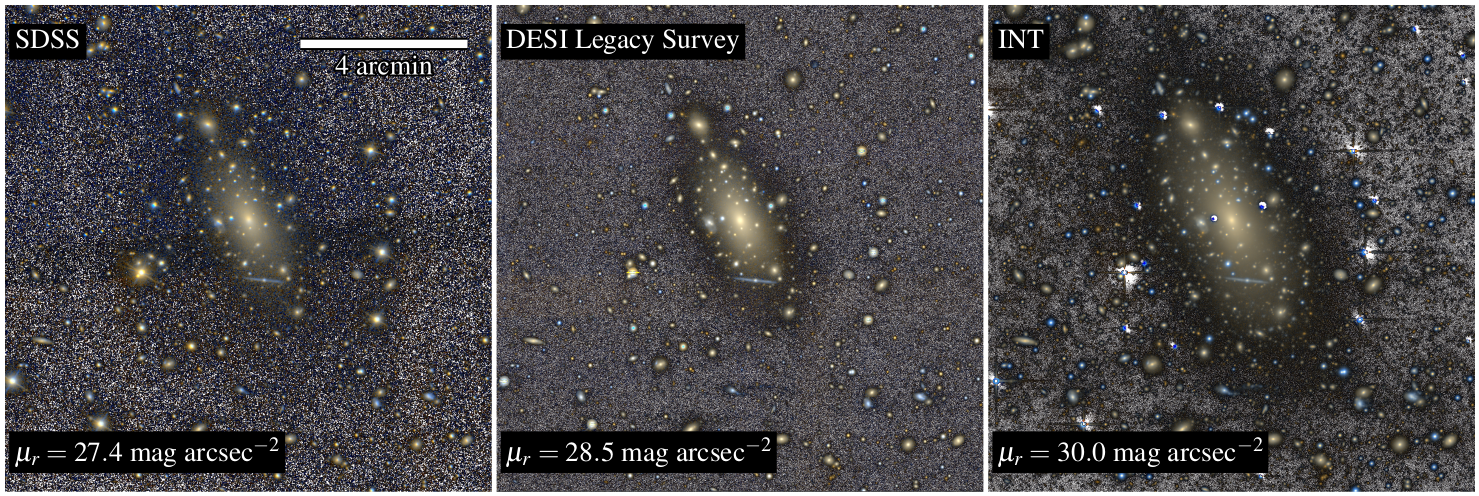}
    \caption{Comparison of colour-composite images of IC~1101 obtained from different imaging surveys. From left to right: SDSS DR16, the DESI Legacy Imaging Surveys DR9 based on DECaLS imaging, and the INT/WFC data presented in this work. In all cases, the colour images were constructed using the $g$- and $r$-bands. For a homogeneous visual comparison, all images were resampled to the plate scale of the INT observations ($0.33''$ pixel$^{-1}$). The quoted values indicate the limiting surface brightness reached in the $r$-band for each dataset, at the $3\sigma$ level within equivalent $10 \times 10$ arcsec$^{2}$ apertures, illustrating the significant gain in depth achieved by the INT observations. A scale bar of 4 arcmin is shown for reference. The INT/WFC image corresponds to our final processed data, after subtraction of the extended haloes of bright foreground stars. The DECaLS image already includes a bright-star treatment performed by the Legacy Survey pipeline (\url{https://github.com/legacysurvey/legacypipe}), while the SDSS image is shown without any additional bright-star subtraction. The effective exposure time ($t_{\rm eff}$) in the $r$-band for the Legacy Survey image is $t_{\rm eff}=3.67$~min, obtained with the NSF Víctor M. Blanco 4-m telescope. In the case of SDSS, the region around IC~1101 is covered by two to three overlapping exposures, resulting in a total $r$-band effective exposure time of $t_{\rm eff} \sim 1.8$--$2.7$~min, acquired with the 2.5-m Sloan Foundation Telescope. The INT/WFC data correspond to a total $r$-band exposure time of $t_{\rm eff}=4.41$~h, highlighting the substantial increase in depth achieved in this work.}
    
    \label{fig:comparison_image}
\end{figure*}

To place the depth of the INT observations in context, Fig.~\ref{fig:comparison_image} presents a direct visual comparison between colour composite images of IC~1101 obtained from the Sloan Digital Sky Survey Data Release~16 \citep[SDSS DR16; ][]{york2000,sdss2020}, the DESI Legacy Imaging Surveys Data Release~9 \citep[Legacy DR9; ][]{dr9legacy2021}, and the INT/WFC data used in this work. All colour images were constructed using the $g$- and $r$-bands, employing the \texttt{astscript-color-faint-gray} algorithm of \citet{infante2024}. This comparison highlights the substantial gain in surface brightness depth achieved by the INT observations, which enables the detection of diffuse LSB structures that are either marginally visible or entirely absent in shallower datasets.

The observational and data processing strategy closely follows the methodology described in \citet{golini2025}, which is similar to the approach outlined in \citet{trujillo2016} and further developed by \citet{zaritsky2024}. In brief, the data were obtained using a dithering pattern with offsets of $\sim4$ arcminutes to improve background estimation and to eliminate detector artifacts and minimize scattered light. The processing pipeline includes a running flat-field correction \citep{saremi2025}, background modelling and subtraction, astrometric alignment, photometric calibration, and final co-addition of the dithered frames into a deep mosaic. Further details on the pipeline are presented in \citet{guerra2025} and \citet{junais2025}.

\section{Methods}
\label{sec:methods}
This section describes the methodology adopted to process the INT/WFC imaging data of IC~1101 in the LSB regime. 
We first detail the construction of an extended PSF tailored to the INT/WFC observations, including source masking, star selection, and the assembly of a radially continuous PSF model. 
We then describe the treatment of scattered light from both foreground stars and IC~1101 itself, combining PSF-based star subtraction with deconvolution techniques based on the wavelet transform \citep{starck2010,carrillo2012}. 

\subsection{Image masking and background treatment}
\label{section:image_masking}

As a first step, an accurate masking of all detected sources in the field is required to prevent contamination in the characterization of the extended PSF, the subtraction of scattered light, the background estimation, and the subsequent extraction of radial profiles. Given the different requirements of these tasks, we adopted a hybrid masking strategy based on two complementary segmentation approaches.

For the construction of the extended PSF and the modelling and subtraction of the scattered light produced by bright stars, we primarily used the tools \texttt{NoiseChisel} and \texttt{Segment} (\citealt{akhlaghi2015}; \citealt{akhlaghi2019}), which are part of the GNU Astronomy Utilities (GNUastro) suite. GNUastro is particularly important for the technical steps associated with PSF construction and stellar-halo subtraction, where a robust identification of both compact and extended sources is needed over a wide range of spatial scales and surface-brightness levels.

For the scientific analysis of IC~1101 itself, in particular the extraction of surface-brightness, colour, and stellar-mass-density profiles, we mainly relied on the \texttt{MTO} (Max-Tree Objects) algorithm (\citealt{teeninga2015,faezi2024}). This method is especially effective at deblending compact sources superimposed on extended emission and in crowded regions, which are common in the environment of massive galaxies such as IC~1101. Therefore, while the GNUastro-based masks were used as the basis for the technical treatment of the images, the MTO masks provided the primary segmentation for the final scientific measurements. For the final extraction of the radial profiles, we constructed a combined mask from the $g$- and $r$-band images, ensuring that all contaminating sources identified in either band are consistently excluded. This approach minimises residual contamination due to colour-dependent detection effects and provides a homogeneous basis for deriving reliable profiles across both filters. A detailed description of both segmentation strategies, together with the adopted parameters and their specific use in the different stages of the analysis, is provided in Appendix~\ref{appendix:segmentation}.

Once the masking was applied, we estimated and corrected for any residual large-scale background remaining after the data reduction. For each band, we computed a circularised radial profile from the masked image, taking the location of IC~1101 as the centre, and identified the outer radial range where the profile becomes flat within the uncertainties, following a procedure similar to that of \citet{pohlen2006}. The residual background level was then defined as the mean surface brightness measured in this radially stable external region and subtracted from the image.

\subsection{Construction of the INT/WFC PSFs} 
\label{sec:INT_PSF}

The PSF encodes the instrumental and atmospheric response to point sources. It plays a central role in studies of the LSB regime, where scattered light from bright objects can dominate the outer regions of galaxies and other diffuse structures, biasing their measured properties \citep[e.g.,][]{uson1991, slater2009, trujillo2016,infante2020, roman2020, montes2021, sedighi2024,golini2025}. In this context, we describe below the methodology adopted to construct the extended PSF for the INT/WFC. The PSF was primarily constructed from the IC~1101 data. However, to characterise its outermost regions, we complemented this with the bright star HD~8648 ($m_{Gaia,G}=7.24$ mag) from the field of NGC~521, as no stars of comparable brightness are present in the IC~1101 field.

The tools developed for these tasks include two custom Python codes. The first, \texttt{LISAN}\protect\footnotemark\footnotetext{\url{https://github.com/CarlosMDLR/LISAN}}
 (Layered Intensity Spread and Analysis for Night-sky structures), is dedicated to the construction and calibration of extended PSFs over wide-field imaging datasets. The second, \texttt{MAHDI}\protect\footnotemark\footnotetext{\url{https://github.com/CarlosMDLR/MAHDI}}
 (Mitigation Algorithm for Halo and Diffuse Illumination), performs scattered light subtraction by modelling and removing the halo of each star using the previously characterised PSF. Both codes are publicly available via their corresponding GitHub repositories and have been successfully applied in previous studies \citep{marrero2026}. The selection of the stars in the image to build the PSF is explained in Appendix \ref{appendix:catalogs_PSF}.

\subsubsection{Radial characterization of the PSF} \label{section:Radial_PSF}

To robustly characterize the PSF, the central parts require unsaturated stars to preserve the PSF’s core structure, while the outskirts need a high signal-to-noise (S/N) ratio, which is only achievable with sufficiently bright stars. However, in the case of the IC~1101 dataset, stars brighter than $m_{Gaia,G}\simeq$14.8 mag suffer from saturation effects (see Fig. \ref{fig:HSM_Mag}). To overcome this limitation, and ensure a consistent PSF model across the entire radial extent, it is necessary to combine stars within different magnitude ranges.

Following a methodology similar to that described in \citet{infante2020} and \citet{sedighi2024}, we adopted a strategy based on dividing the stellar sample into four distinct magnitude intervals, each optimized to characterize a specific radial regime of the PSF. 

To define appropriate magnitude intervals for the construction of the PSF, we first consider two key constraints. On one hand, the PSF must be traced across a broad dynamic range to characterize both its core and extended wings. On the other hand, instrumental effects such as internal reflections must be carefully handled. In particular, the WFC on the INT exhibits prominent internal reflections arising from the secondary mirror, especially in the presence of bright stars. These reflections have a complex morphology that varies across the detector and become more symmetric when the star is located near the optical axis. Since internal reflections are an intrinsic component of the PSF, they are present for stars of all magnitudes, although they are only detectable at large radii for sufficiently bright sources. 

For this reason, dividing the PSF construction into separate magnitude and radial intervals helps to optimise the signal-to-noise ratio across different regions: the brightest stars constrain the outer PSF wings, where the signal from fainter stars falls below the detection limit, while fainter stars provide a more reliable characterization of the inner regions, where the profiles are less affected by saturation and non-linear effects.

However, if we define large intervals, stars with significantly different brightness levels leads to heterogeneous profiles that compromise the precision of the stacked PSF, particularly in the low-S/N outskirts \citep{infante2020,sedighi2024}. To mitigate this effect, we restrict each magnitude interval to a maximum width of one magnitude (i.e. the difference between the faintest and brightest star within each bin), ensuring homogeneous S/N and consistent radial behaviour within each subset. The adopted bins are defined along the point-like source branch identified in Fig.~\ref{fig:HSM_Mag}, and correspond to: $19 \leq m_{Gaia,G} \leq 20$ (Inner), $16.9 \leq m_{Gaia,G} \leq 17.9$ (Subintermediate), $14.8 \leq m_{Gaia,G} \leq 15.8$ (Intermediate), and $m_{Gaia,G} < 14.8$ (Outer).

The three one-magnitude intervals spanning $14.8 \leq m_{Gaia,G} \leq 20$ are composed of unsaturated stars and are used to characterise the PSF core and its intermediate radii. Rather than adopting the brightest unsaturated source inside the point-like source branch as the boundary, we conservatively define the transition slightly below the onset of saturation along this branch, thereby avoiding stars that may already exhibit non-linear detector effects. Stars brighter than $m_{Gaia,G} = 14.8$ populate the Outer bin and are employed to build the extended PSF wings.

To minimize contamination from edge effects that may distort the photometry of stars located near the image boundaries, a spatial selection is applied to the stars used in the three internal PSF regions (labelled Inner, Subintermediate, and Intermediate). Specifically, we restrict the selection to stars located within circular apertures centred on the galaxy, with maximum radii of 10\arcmin, 12\arcmin, and 14\arcmin\ for the three regions, respectively. These limits are designed also to consider the decreasing number of stars available toward brighter magnitude bins.

Once the stars have been selected, their centroids are refined with sub-pixel accuracy to ensure precise alignment during stacking. Image cutouts centred on the \textit{Gaia} DR3 coordinates are extracted and collapsed along the $x$ and $y$ axes to obtain one-dimensional flux profiles. These profiles are fitted with a Lorentzian function, which provides an adequate description of stellar cores with extended wings in ground-based data \citep{howell2006}. Although alternative forms such as the Moffat profile are commonly used \citep[e.g.,][]{trujillo2001}, tests performed with both functions show a negligible impact on the centroid determination: the median absolute difference between Lorentzian and Moffat centres is $0.005 \pm 0.005$ arcsec. 

Moreover, the offset between the original \textit{Gaia} coordinates and the refined Lorentzian centroids has a median value of $0.08 \pm 0.07$ arcsec. Given that this difference is well below the pixel scale, we adopt the \textit{Gaia} centres for saturated stars and for the subsequent star-subtraction procedure, as this approximation does not introduce any significant systematic uncertainty.

\begin{figure}[t]
    \centering
    \includegraphics[width =\columnwidth]{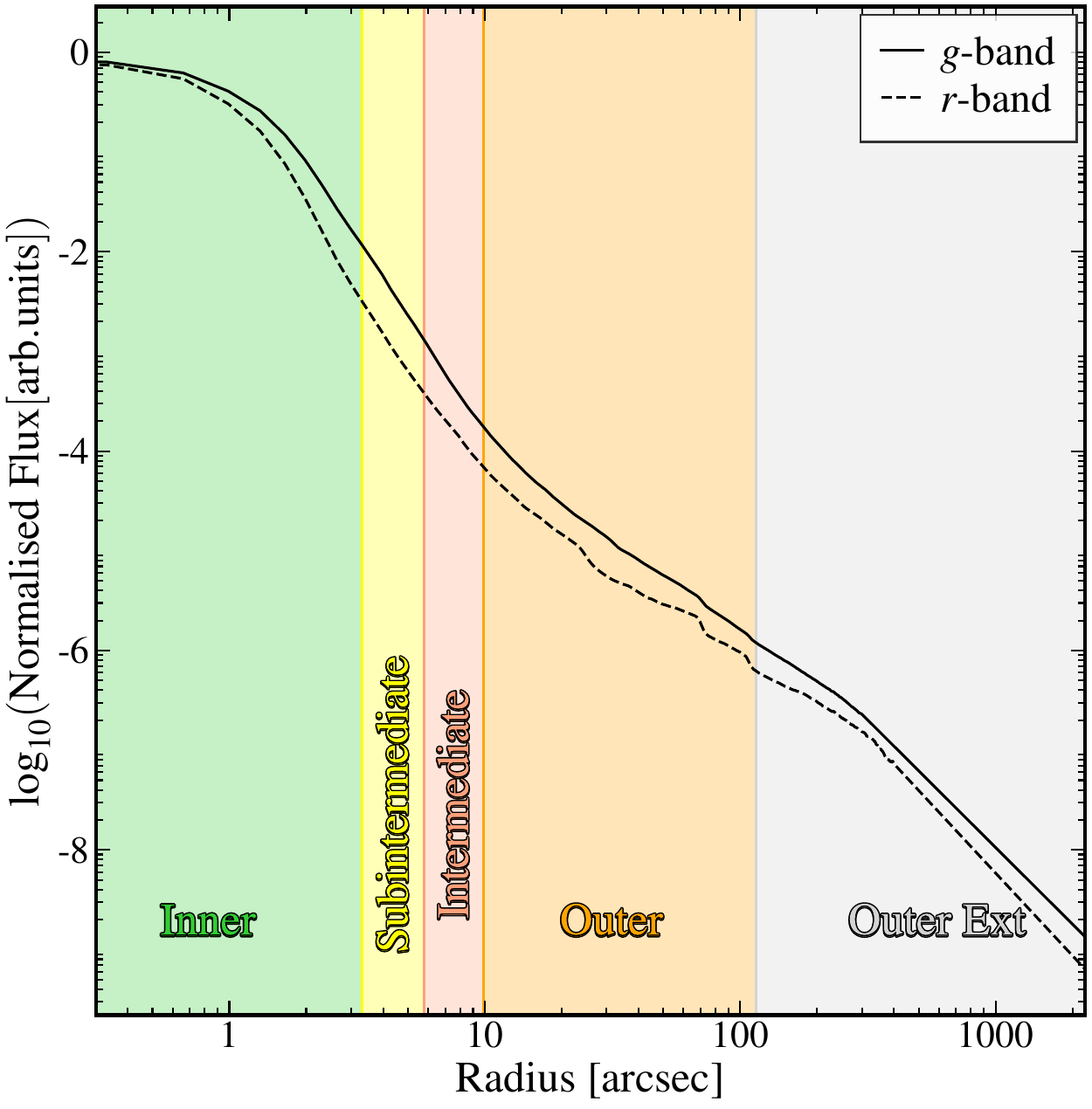}
    \caption{Radial profile of the extended PSF in the $g$- and $r$-bands, normalized to unity, by dividing for the total flux. The different shaded regions highlight the radial intervals used to construct the PSF: Inner, Subintermediate, Intermediate, Outer and Outer Ext. These regions correspond to different magnitude regimes used to characterize the PSF at various spatial scales.}
    \label{fig:psf}
\end{figure}

Then, the radial range referred to as Outer was further subdivided into 1-magnitude-wide intervals up to the magnitude of the brightest star in the image. From these intervals, we selected the brightest magnitude bin, which provides the highest S/N in the outer region, to construct the Outer section of the PSF.

Once the automatic selection process is complete, the algorithm displays cutouts of all candidate stars for each magnitude range, allowing for visual inspection. This additional step enables a final quality check, during which stars that are not suitable for the PSF construction can be manually rejected. In the $g$ band, the final number of stars retained after visual inspection was 53 for the inner PSF region, 8 for the sub-intermediate region, 9 for the intermediate region, and 10 for the outer region.

For all PSF sections the selected stars were combined using a sigma-clipped mean with a $2\sigma$ threshold. This procedure efficiently removes possible artefacts while preserving the core and wings of each stellar profile.

Due to the lack of sufficiently bright stars in the immediate surroundings of IC~1101, the outermost region (Outer Ext) was characterized using an external calibration star: HD~8648, a DR3 source with $m_{Gaia,G} = 7.24$~mag. This star was observed with the same instrument (INT/WFC) in the field of the galaxy NGC~521, providing a robust constraint on the PSF wings. The data consist of 135 individual exposures of 200 s in the $g$-band, corresponding to a total exposure time of 7.5 hours, ensuring a high signal-to-noise characterization of the outer PSF. Although obtained under similar observational conditions, we note that differences in observing epochs may introduce some uncertainty, as the outer PSF can vary with time (e.g. due to changes in mirror cleanliness), which should be considered when interpreting this component.

It is important to note that the resulting PSF models may still be marginally broader than the true instrumental PSF. This is mainly because the PSF shape, especially its degree of circularity, can vary across the field of view. Although the sigma-clipped stacking statistically mitigates these spatial variations, some level of broadening is unavoidable when combining stars from different detector regions.

\subsubsection{Assembling the final PSF}\label{sec:uniting_parts}

Once all sections have been constructed, a significant challenge remains: combining the different parts into a coherent whole. The stars in each section have been normalized using a reference ring to align them at the same level. However, differences in brightness persist across the sections. To ensure a smooth transition between them, we adopt the approach outlined in \citet{infante2020}, joining the profiles at points where the S/N is similar. This process is carried out sequentially, starting from the outermost region and moving inward. The joining regions and the extrapolated behaviour of the PSFs at large distances are described in Appendix \ref{appendix:regions_PSF}.

The final normalized PSFs are shown in Fig.~\ref{fig:psf}. Both $g$ and $r$ profiles exhibit a smooth and continuous behaviour across all radial ranges, with consistent shapes and well-matched transitions between the inner and outer regions. The different PSF regions discussed above are represented in the figure, illustrating the full radial construction of the model. Small bumps are visible between $\sim10''$ and $\sim100''$, particularly in the $r$-band, which arise from residual reflections of the telescope’s secondary mirror. Despite these minor features, the extended power-law tails ensure accurate modelling of scattered light over the full spatial extent of IC~1101.

\subsection{Treatment of scattered light from stars and the galaxy itself}\label{sec:treat_scatter}

\begin{figure*}[t]
    \centering
    \includegraphics[width =0.7\linewidth]{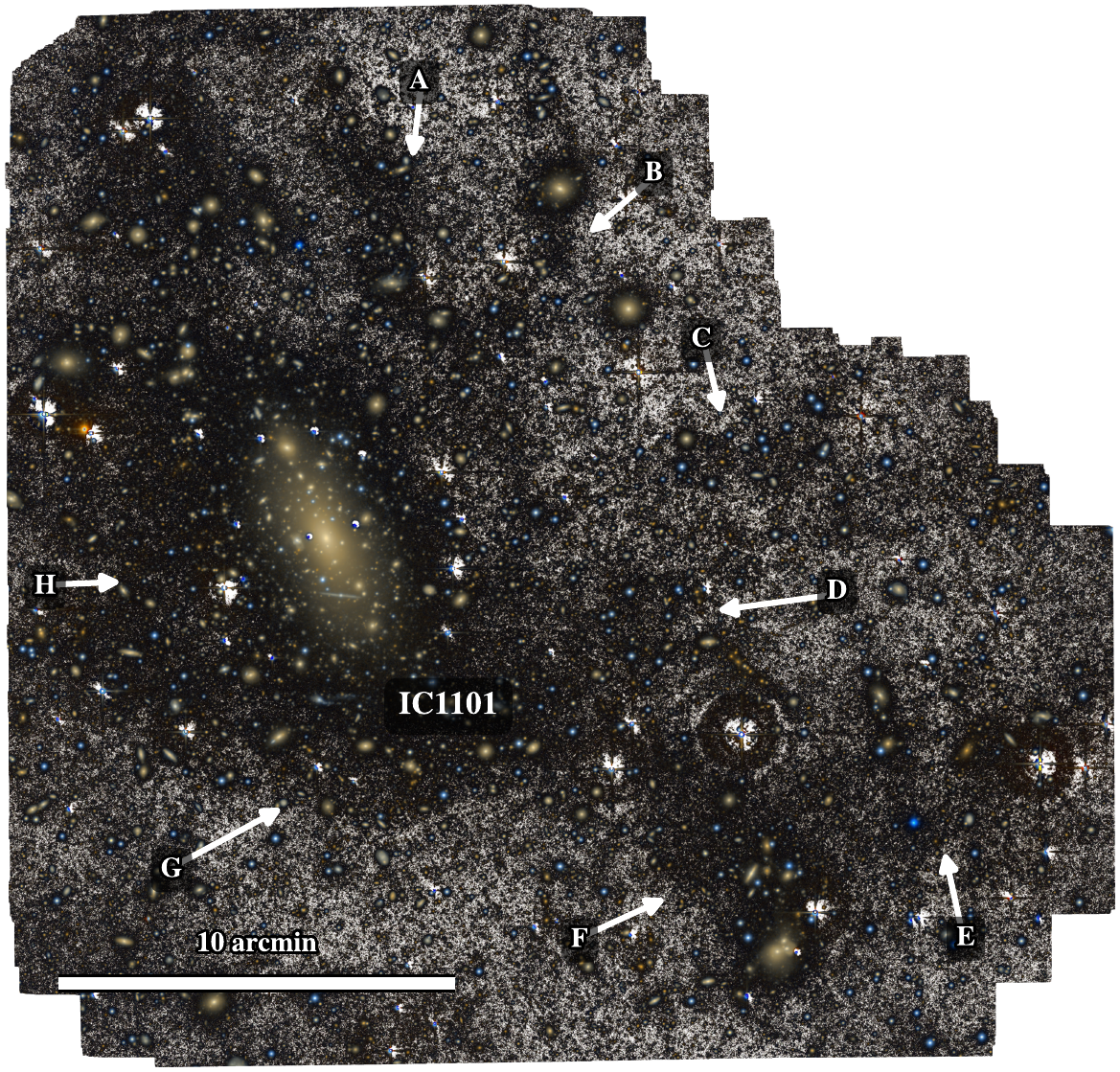}
    \caption{Colour composite image of IC~1101, based on the INT/WFC data presented in Sect.~\ref{sec:data}, after applying the star subtraction algorithm. The image covers a field of view of approximately $28.1' \times 26.9'$. The image was created using the Sloan $g$- and $r$-bands and the \texttt{astscript-color-faint-gray} algorithm described in \citet{infante2024}. For the visualization of the background and diffuse light, the $r$-band was used as the base greyscale layer. White arrows and labels from A to H indicate the LSB features discussed in the text. In the greyscale representation, lighter pixels correspond to sky-dominated regions, while darker pixels trace brighter areas with higher signal.}

    \label{fig:full_color_image}
\end{figure*}

\subsubsection{Subtracting stars}
\label{sec:subtract_stars}

The scattered field represents the diffuse background generated by the extended PSF wings of stars, whose light can spread over several arcminutes and contaminate the LSB regions of the image \citep[e.g.,][]{capaccioli1983, sandin2014, trujillo2016, infante2020, sedighi2024}. After properly characterizing the PSF in both bands, the stars can be subtracted from the images in order to mitigate this effect. For this purpose, we modelled and removed the scattered light field produced by all stars in the image with $m_{\mathrm{Gaia},G} < 16$~mag in both the $g$- and $r$-filters. This magnitude limit was adopted because fainter stars do not contribute significantly to the large-scale scattered light at the galactocentric distances relevant for our analysis, as their extended PSF wings fall below the surface-brightness levels of interest \citep[see also][]{sedighi2024}. In total, 250 stars were selected in both the $g$- and $r$-bands for this correction. A visualization of the resulting scattered light map is provided in Appendix~\ref{appendix:scatter light map}.

The modelling of each star involves two main steps: determining its centre and scaling the PSF model to match the stellar flux. Stellar centroids are adopted from the \textit{Gaia} DR3 coordinates, while the photometric scaling is obtained by comparing the observed radial profile of each star with the reference PSF profile. To ensure that this normalization is performed in a regime that is simultaneously free of saturation (inner pixels of bright stars) and robust against noise (outer radii of faint stars), we define a magnitude-dependent annular region where the PSF scaling factor is computed.

We compute a scaling factor for each star, defined as the 2$\sigma$-clipped mean ratio between the observed surface brightness profile and the PSF model. With the stellar centroids from \textit{Gaia} DR3 and the corresponding flux normalization determined, we scale and subtract the stars from the image. To do so, we build a hierarchical scattered light map by adding and subtracting the contribution of each star sequentially, starting from the brightest and proceeding to progressively fainter sources. In practice, the brightest star is modelled and removed first, followed by the next brightest, and so on. This iterative approach is necessary because the extended halos of bright stars can significantly affect the scaling and subtraction of nearby, fainter sources.

Fig.~\ref{fig:full_color_image} illustrates the effect of this correction on the final colour-composite image of IC~1101, showing a clear removal of the diffuse halos produced by foreground stars, particularly around the brightest ones. Once the scattered light contamination is removed, several LSB regions become visible across the field. In the upper and right-hand sides of the image, faint filamentary structures (labelled A, B, and C) extend toward the cluster centre. In contrast, broader and more diffuse features located to the left and below the galaxy (labelled H, G, and D) suggest a different origin or dynamical configuration. Additional substructures, marked as E and F, also appear to trace ongoing or recent accretion events. These features will be analysed in detail in Section~\ref{sec:discussion}. The overall background becomes smoother and more uniform than in the original image, and the faint outskirts of the galaxy are more clearly revealed. Further details on the impact of the star--subtraction procedure are presented in Appendix~\ref{appendix:scatter light map}, where a more detailed comparison between the images before and after star subtraction is shown, together with the corresponding scattered--light maps. This appendix illustrates the effect that scattered light can have on deep imaging data and emphasizes the importance of properly modelling and removing this component to reliably unveil faint LSB structures.

\subsubsection{Wavelet-Based Scattered Light Mitigation}\label{sec:Wavelet}

To mitigate scattered light arising from the extended stellar emission of IC~1101 and from nearby galaxies in its local environment, we applied a wavelet-based decomposition technique \citep{starck2010,carrillo2012}. This approach suppresses contamination from PSF wings and removes diffuse emission from extended sources that are not well represented by a point-source PSF. Further technical details of the adopted wavelet framework, including the regularisation scheme and optimisation strategy, are provided in Appendix~\ref{sec:wave_framework}.

To minimise the introduction of high-frequency noise and the potential degradation of faint LSB features associated with direct deconvolution, we adopt a model-based PSF convolution technique following \citet{golini2025}. This approach allows us to isolate and remove the contribution of the PSF from extended stellar emission in a controlled manner. The correction is applied to the full science image after foreground star subtraction (Sect.~\ref{sec:subtract_stars}). Further details of this procedure are provided in Appendix~\ref{appendix:sb_deconvolution}, while a schematic illustration of the method is shown in Fig.~2 of \citet{golini2025}.

\section{Radial structure of IC~1101}
\label{sec:radial_structure}

\begin{figure*}[t]
    \centering
    \includegraphics[width=0.7\textwidth]{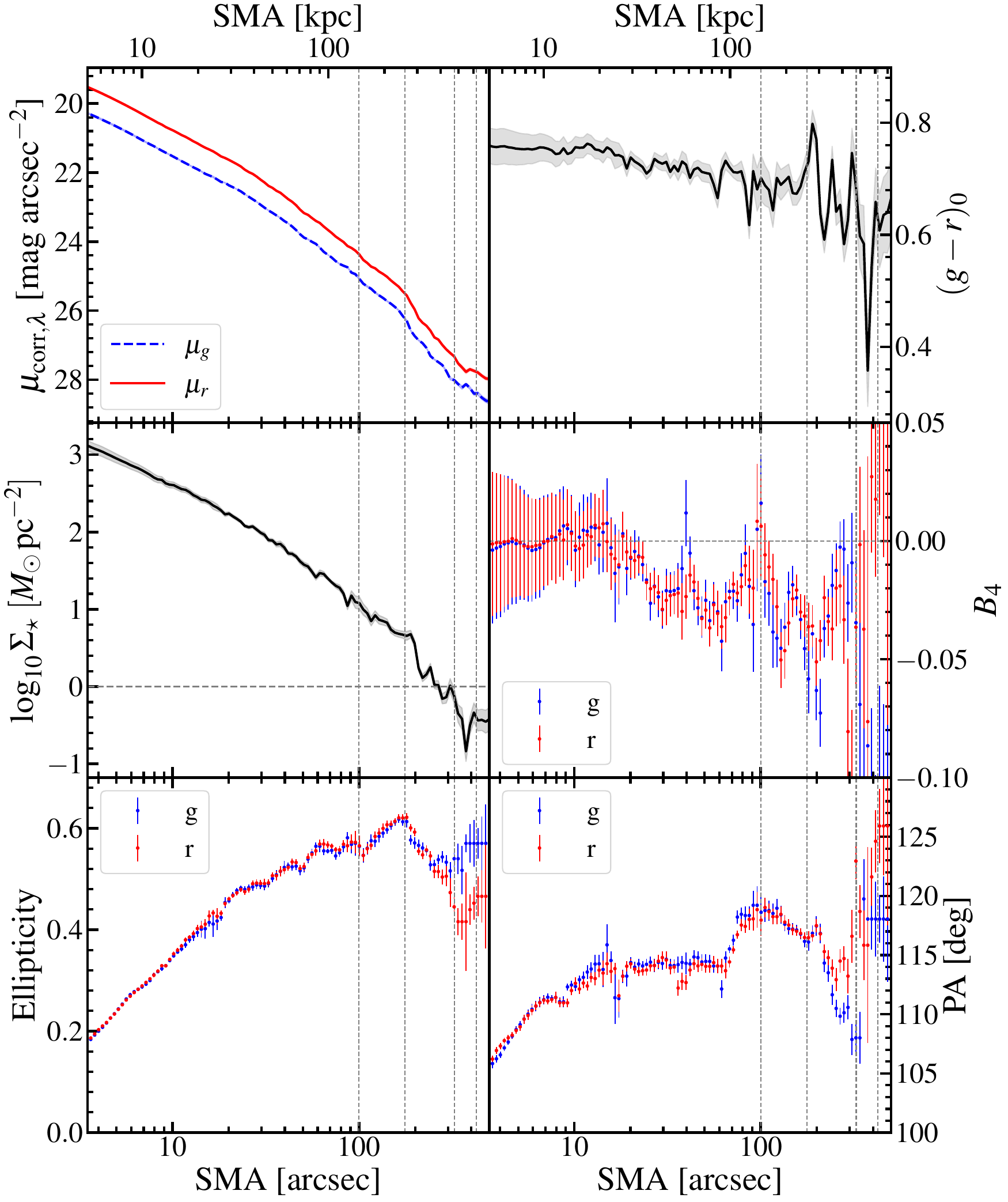}
    \caption{
    Radial profiles of IC~1101 derived using \texttt{photutils}, allowing the ellipticity and position angle to vary freely with radius. 
    The panels show the surface-brightness profiles in the $g$ and $r$ bands, the $(g-r)_0$ colour profile, the stellar mass surface density profile, the $B_4$ coefficient, the ellipticity, and the position angle. 
    Vertical dashed lines mark the characteristic radii discussed in the text. 
    The most prominent transition occurs at $\sim260$ kpc, where both the ellipticity and position angle show significant changes, accompanied by coherent variations in colour, surface brightness, and stellar mass density. 
    This feature is adopted as the fiducial edge radius of the BCG.
    }
    \label{fig:photutils_profiles_free}
\end{figure*}

In this section, we analyse the radial structure of IC~1101 by combining three complementary diagnostics. 
First, we derive azimuthally averaged elliptical profiles in which both the position angle (PA) and ellipticity are allowed to vary freely with radius. 
This provides a non-parametric description of the galaxy structure and allows us to identify the main radial transitions without imposing a fixed geometry. 
Second, we use wedge-based profiles aligned with the major axis of IC~1101 to trace the low surface-brightness emission to larger galactocentric distances. 
Third, we examine two-dimensional maps of surface brightness, colour, and stellar mass surface density to assess whether the one-dimensional transitions correspond to coherent spatial structures.

\begin{figure*}[t]
    \centering
    \includegraphics[width =\linewidth]{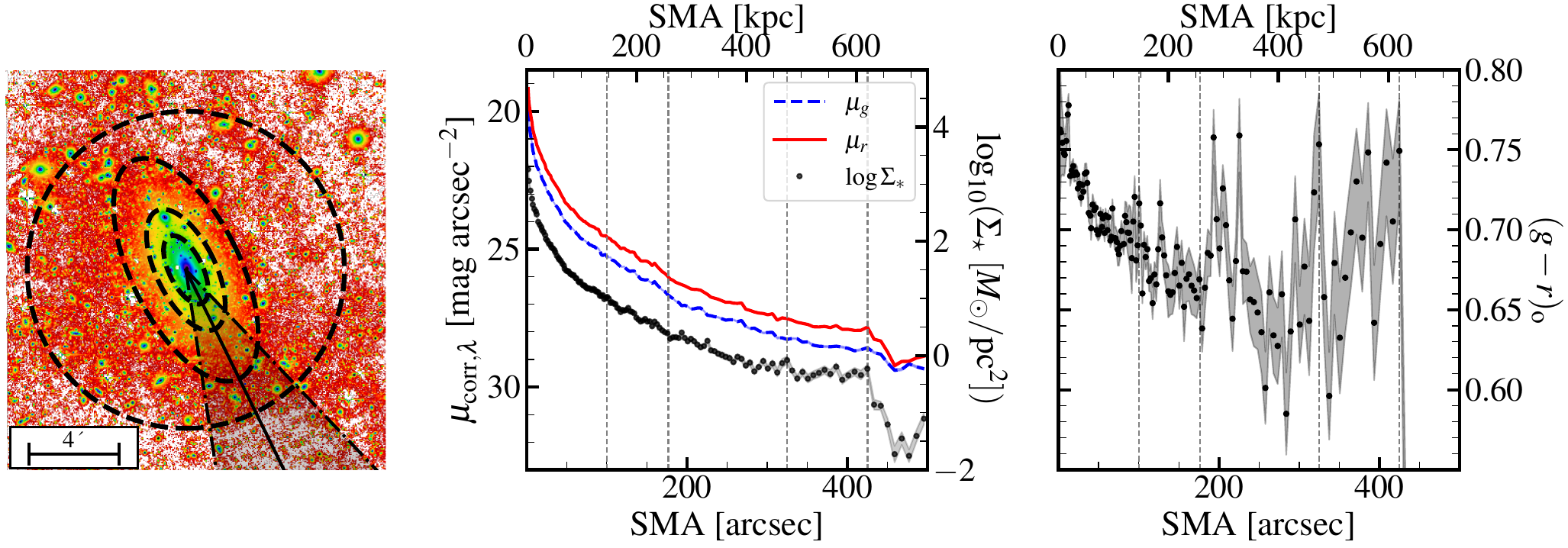}
    \caption{%
    Left panel: PSF deconvolved image of IC~1101 in the $r$ band. A $4'$ scale bar is shown, together with the wedge used to extract the major-axis profiles. Four characteristic radii, the first two based on the analysis of the profiles shown in Fig.~\ref{fig:photutils_profiles_free}, are marked: $\sim146$ kpc (100\arcsec), $\sim260$ kpc (180\arcsec), $\sim475$ kpc (325\arcsec), and $\sim620$ kpc (425\arcsec). 
    Middle panel: Surface-brightness profiles in the $g$ and $r$ bands, together with the stellar mass surface density profile. 
    Right panel: Extinction-corrected colour profile, $(g-r)_0$.
    }
    \label{fig:profiles}
\end{figure*}

The radial profiles, derived using the \texttt{photutils} package \citep{jedr1987, bradley2019}, are shown in Fig.~\ref{fig:photutils_profiles_free}. 
They include the surface-brightness profiles in the $g$ and $r$ bands, the extinction-corrected $(g-r)_0$ colour profile, the stellar mass surface density profile, the fourth-order Fourier coefficient $B_4$, the ellipticity, and the PA. The associated uncertainties were computed following the procedure described in \citet[][Eqs.~1--3]{golini2024}. In this approach, the uncertainty at each radial distance is estimated from the dispersion of the unmasked pixel values within the corresponding elliptical annulus, together with the dispersion measured in the background regions. The final error assigned to each point therefore represents the uncertainty on the mean intensity measured over the annulus, after propagating the contribution from the background determination. The surface-brightness profiles were corrected for both Galactic extinction and cosmological surface-brightness dimming. Hence, the fully corrected surface-brightness profiles in each band are thus defined as

\begin{equation}
\mu_{\mathrm{corr},\lambda} = \mu_{\mathrm{obs},\lambda} - A_{\lambda} - 7.5 \log_{10}(1+z),
\end{equation}

\noindent where $\mu_{\mathrm{obs},\lambda}$ is the observed profile, $A_{\lambda}$ is the Galactic extinction in the corresponding band. The last term accounts for cosmological dimming that affects the surface brightness as $(1+z)^3$  \citep[e.g.,][]{giavalisco1996,ribeiro2016,buitrago2024}. The extinction values are taken from NASA NED Foreground Galactic Extinctions \citep{schlafly2011}, with $A_g = 0.131$ and $A_r = 0.090$.

The stellar mass surface density profiles are derived using a standard approach based on the galaxy’s surface brightness and colour information. In particular, we adopt the following expression \citep[see ][]{bakos2008}:

\begin{equation}
    \log_{10} \Sigma_{\star} = \log_{10} (M/L)_{\lambda} - 0.4\ (\mu_{\mathrm{corr},\lambda} - m_{\odot,\lambda}^{\mathrm{abs}}) + 8.629,
\end{equation}

\noindent where $\mu_{\mathrm{corr},\lambda}$ denotes the corrected surface brightness in band $\lambda$, $m_{\odot,\lambda}^{\mathrm{abs}}$ is the absolute magnitude of the Sun in that band, and $(M/L)_{\lambda}$ is the corresponding mass-to-light ratio. The latter is inferred from a colour–$M/L$ relation of the form

\begin{equation}
    \log_{10} (M/L)_{\lambda} = a_{\lambda} + b_{\lambda} \times \mathrm{colour},
\end{equation}

\noindent where the coefficients $a_{\lambda}$ and $b_{\lambda}$ depend on the adopted band and stellar population synthesis model.

In this work, we select the $g$-band as the reference owing to its greater depth. The colour term is defined as $(g - r)$, and we adopt $a_g = -0.984$ and $b_g = 2.029$, following \citet{roediger2015}, that correspond to a Chabrier initial mass function (IMF; \citealt{chabrier2003}). The absolute magnitude of the Sun in the $g$-band is taken to be $m_{\odot,g}^{\mathrm{abs}} = 5.11$, as reported by \citet{willmer2018}.

By looking at Fig. \ref{fig:photutils_profiles_free}, a first transition can be visually identified at $R \sim 100''$ ($\sim146$~kpc). The signatures at this location are seen in the ellipticity and PA profiles, which show noticeable variations with respect to the inner regions. The colour profile may also suggest a mild change towards a flatter behaviour. As discussed below, the surface-brightness and stellar-mass-density profiles do not show their strongest variation at this radius. 
This feature may mark the onset of a more mixed stellar population or an internal structural change within the BCG. 
However, the surface-brightness and stellar mass surface density profiles do not show their strongest variation at this location. 
We therefore interpret this radius as an internal transition rather than as the outer edge of the BCG.

A more prominent transition is found at $R \sim 180''$ ($\sim$260 kpc). 
At this radius, the ellipticity decreases and the PA changes significantly, indicating a transition towards a rounder and more circularized configuration. 
This behaviour is accompanied by a clear variation in the colour profile and by changes in the slopes of the surface-brightness and stellar mass surface density profiles. 
The simultaneous presence of this feature in the structural parameters and in the photometric profiles makes it the most robust radial transition identified in IC~1101.

We adopt the transition at $\sim260$ kpc as our fiducial estimate of the edge radius of the BCG, $R_{\mathrm{edge}}$.
It corresponds to the innermost radius at which the main stellar body of IC~1101 shows a strong and coherent change in its structural and photometric properties. 
In early-type galaxies, the edge is not expected to appear as a sharp truncation, but rather as a smooth transition in the slope of the surface-brightness and stellar mass density profiles, often accompanied by changes in colour and geometry \citep{trujillo2020,chamba2022}. 
In this framework, $R_{\mathrm{edge}}$ traces the boundary between the main stellar body of the galaxy and the outer envelope assembled through accretion and merging events. 
The feature at $\sim260$ kpc matches this phenomenology and provides the clearest observational evidence for such a boundary in IC~1101.

After identifying the main structural transitions from the isophote analysis, we extract wedge-based profiles along the major axis of IC~1101 in order to follow the low surface-brightness emission to larger radii. 
This approach maximizes the signal in the direction where the extended emission is most prominent and avoids mixing regions with different structural behaviour. At $\sim260$~kpc, the profiles yield an axis-ratio of $q = 0.38 \pm 0.02$ and a position angle of $PA = 116.1^{\circ} \pm 3.0^{\circ}$. 
For the wedge-based analysis, we adopted this position angle as the reference orientation of the major axis. 
However, we fixed the axis ratio to $q=1$ when extracting the profiles with the \texttt{astscript-radial-profile} task from the \texttt{GNUastro} suite \citep{infante2024radial}. 
This choice was made because the wedge analysis is intended to explore the outermost regions of the system, where the profiles show a sharp decrease in ellipticity and the stellar distribution becomes progressively rounder and less well described by a single elliptical geometry. The resulting profiles are shown in Fig.~\ref{fig:profiles}, together with the corresponding $(g-r)_0$ colour profile and stellar mass surface density profile. 
The same characteristic radii discussed throughout this section are overlaid on the profiles and on the image.

The wedge-based profiles confirm the radial features identified from the isophote analysis and extend the analysis to lower surface-brightness levels. 
The transition at $\sim146$ kpc is mainly visible as a change in the colour gradient, consistent with the behaviour observed in the isophote analysis. 
The feature at $\sim260$ kpc is detected as a coherent variation in surface brightness, colour, and stellar mass density, in agreement with the structural transition seen in the ellipticity and PA profiles. 
This supports its interpretation as the fiducial edge radius of the BCG.

At larger radii, the wedge-based profiles reveal two additional features. 
The first one appears at $R \sim 325''$ ($\sim475$ kpc), where the surface-brightness and stellar mass surface density profiles show a further change in behaviour. 
In the image, this radius also corresponds to the region where the regular elliptical morphology traced at smaller radii starts to dissolve into a more diffuse and asymmetric low surface-brightness component. 
This transition could therefore be interpreted as a possible structural boundary of the extended stellar distribution. 
However, in dynamically evolved cluster environments, the outer envelope of the BCG and the ICL are expected to be spatially coupled and may share similar geometric properties \citep{dressler1978,kluge2021}. 
For this reason, the interpretation of this feature as the edge of the BCG alone is less straightforward. 
We therefore interpret it as a transition within the extended BCG+ICL component.

A second outer feature is observed at $R \sim 425''$ ($\sim620$ kpc). 
This radius approximately marks the outer limit of the coherent low surface-brightness emission detected around IC~1101, beyond which the profiles become increasingly irregular and dominated by low S/N fluctuations. 
We therefore interpret this radius as the boundary between the material accreted and the infalling material, the one that has not had time to yet mix with the central BCG+ICL structure.

\begin{figure}[h!]
    \centering
    \includegraphics[width =0.85\columnwidth]{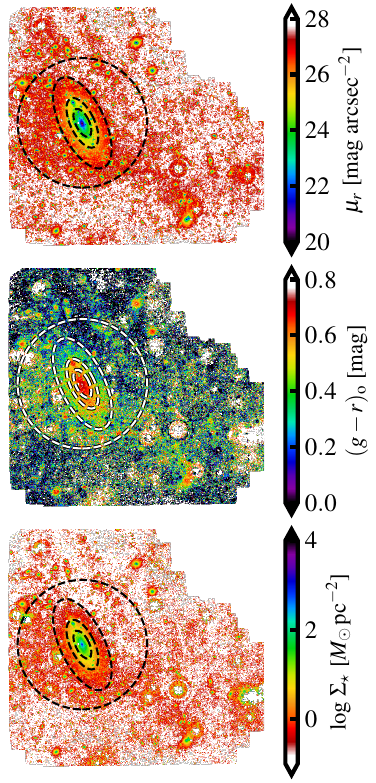}
    \caption{
    Two-dimensional maps of IC~1101 derived from the PSF deconvolved INT/WFC imaging after correcting for Galactic extinction and cosmological surface-brightness dimming. 
    Top panel: $r$-band surface brightness map, $\mu_r$, in units of mag~arcsec$^{-2}$. 
    Middle panel: extinction-corrected colour map $(g-r)_0$. 
    Bottom panel: stellar mass surface density map, expressed as $\log\,\Sigma_\star$ in units of $M_\odot\,\mathrm{pc}^{-2}$, obtained from the surface brightness and colour information. 
    All maps are shown at a spatial resolution corresponding to 2.64 arcsec per pixel. The colour bars indicate the corresponding physical scales for each quantity. The ellipses and the circular aperture overlaid on the maps correspond to those used in Fig.~\ref{fig:profiles}, allowing a direct comparison between the two-dimensional distributions and the extracted radial profiles.
    }
    \label{fig:maps}
\end{figure}

The two-dimensional maps shown in Fig.~\ref{fig:maps} support this interpretation. 
Within $R_{\mathrm{edge}}$, IC~1101 shows a relatively regular and elliptical stellar distribution. 
Beyond this radius, the light distribution becomes progressively more diffuse and less symmetric, while the stellar mass surface density shows a flatter and more extended configuration. 
The colour map also reveals spatial variations associated with the same radial range, consistent with the changes observed in the one-dimensional profiles. 
At larger radii, the maps show that the structures between $\sim475$ and $\sim620$ kpc are more irregular than the inner galaxy body, supporting their interpretation as part of the diffuse BCG+ICL envelope.

\begin{figure*}[h!]
    \centering
    \includegraphics[width =\textwidth]{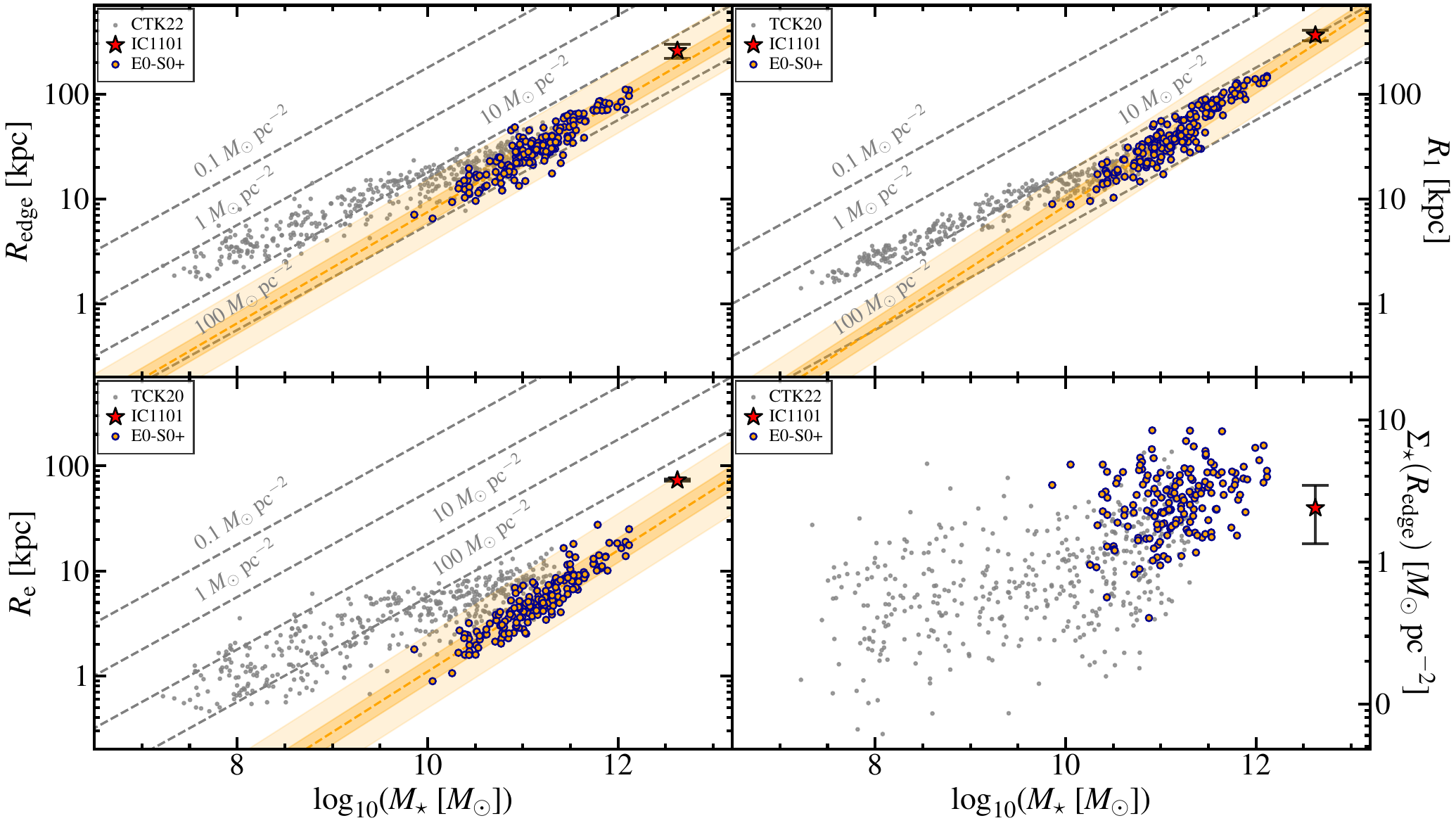}
    \caption{
    Comparison between IC~1101 and galaxies from CTK22 \citep[][]{chamba2022} and TCK20 \citep[][]{trujillo2020} in four structural scaling relations. 
    Top-left: stellar mass versus edge radius ($R_{\mathrm{edge}}$), compared to galaxies from CTK22. 
    Top-right: stellar mass versus $R_1$, using galaxies from TCK20. 
    Bottom-left: stellar mass versus effective radius, $R_{\textrm{e}}$, from TCK20. 
    Bottom-right: stellar mass surface density at $R_{\mathrm{edge}}$, $\Sigma_*(R_{\mathrm{edge}})$, as a function of stellar mass. 
    IC~1101 is shown with a red star marker, while late-type galaxies from TCK20 and CTK22 are shown as grey dots. 
    Early-type galaxies from TCK20 and CTK22 (E0--S0$^{+}$) are shown as orange dots. 
    The coloured shaded regions represent the uncertainty of the linear fit to the E0--S0$^+$ galaxies, with the darker band corresponding to the $1\sigma$ interval and the lighter band to the $3\sigma$ interval of the fit. 
    We include dashed lines indicating constant projected stellar mass surface density of, from top to bottom, 0.1, 1, 10, and 100 $M_\odot$ pc$^{-2}$.
    }
    \label{fig:size_comparison}
\end{figure*}

\begin{figure}[h!]
    \centering
    \includegraphics[width =\columnwidth]{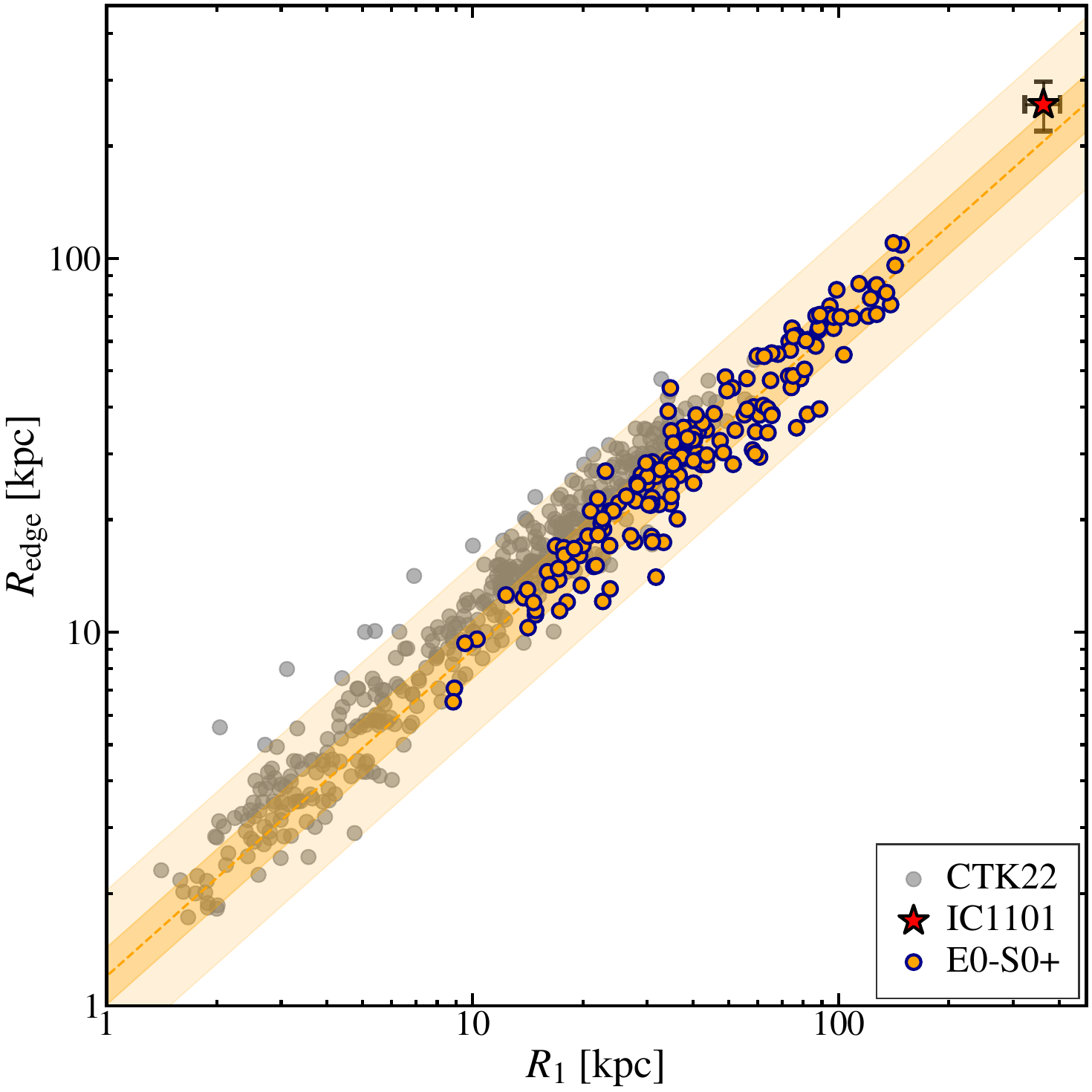}
    \caption{
    Relation between $R_1$ and $R_{\mathrm{edge}}$. 
    Grey points correspond to the CTK22 \citep[][]{chamba2022} sample, while coloured circles denote early-type systems (E0--S0$^+$) from the same sample. 
    IC~1101 is highlighted with a red star. 
    The dashed line shows the best-fitting relation derived for the E0--S0$^+$ systems. 
    The coloured shaded regions represent the uncertainty of this fit, with the darker band corresponding to the $1\sigma$ interval and the lighter band to the $3\sigma$ interval.
    }
    \label{fig:r1vsredge}
\end{figure}

\section{Size and mass measurements}
\label{sec:size_comparison}

To place the spatial extent of IC~1101 in the context of previous observational studies, we complement the edge radius defined in Sect.~\ref{sec:radial_structure} with additional size tracers. In particular, we consider the effective radius, $R_{\mathrm{e}}$, and the $R_1$ radius, which are commonly employed to quantify galaxy sizes and establish scaling relations. While $R_{\mathrm{e}}$ traces the radius enclosing half of the total light \citep[e.g.,][]{vaucouleurs1948, graham2005a} and is therefore sensitive to the inner light distribution and profile shape, $R_1$ \citep[motivated by the gas density star formation threshold in present-day galaxies;][]{trujillo2020} is defined as the radial location where the stellar mass surface density reaches $\Sigma_\star = 1\,M_\odot\,\mathrm{pc}^{-2}$. The $R_1$ metric is now widely adopted in both observational and simulation-based studies \citep[e.g.,][]{chamba2022,arjona2025,claudio2025,marrero2026,ludlow2026}.

Using the wedge-radial profiles derived in this work, we measure $R_1 = 362 \pm 38$~kpc and an effective radius of $R_{\mathrm{e}} = 73 \pm 2$~kpc. 
The $R_1$ radius is comparable to the independently identified edge radius, $R_{\mathrm{edge}} = 260 \pm 38$~kpc. 
At this location, the stellar mass surface density is $\Sigma_\star(R_{\mathrm{edge}}) = 2.40 \pm 1.05\ M_\odot\ \mathrm{pc}^{-2}$. 
The quoted uncertainties in the size parameters follow the methodology adopted in \citet{trujillo2020} and \citet{chamba2022}. 
In particular, we account for the impact of the background determination by adding and subtracting the local background uncertainty from the surface-brightness profiles, and then recomputing $R_1$, $R_{\mathrm{edge}}$, and $R_{\mathrm{e}}$ in each realization. 
This procedure captures the sensitivity of faint isophotal and mass-density thresholds to small background variations, which can significantly shift the inferred sizes in the LSB regime.

For consistency with previous works and to enable a direct comparison with the scaling relations of \citet{trujillo2020} and \citet{chamba2022}, we adopt as representative stellar mass of IC~1101 the mass enclosed within the radius at which the wedge-based $g$-band surface-brightness profile reaches $\mu_g = 29$~mag~arcsec$^{-2}$, following the same criterion used in those studies. This surface-brightness level is reached at $R(\mu_g=29)\sim 652$~kpc. Using this definition, we measure a stellar mass of $(4.2 \pm 0.1)\times10^{12},M_\odot$, which is the value used to place IC~1101 in the mass-size relations shown in Fig.~\ref{fig:size_comparison}.

For reference, the stellar mass enclosed within the inner transition at $\sim146$~kpc is $(2.662\pm 0.004)\times10^{12}\,M_\odot$, while the stellar mass within the edge radius of the BCG is $(3.36 \pm 0.01)\times10^{12}\,M_\odot$. The mass within $R_1$ is $(3.69 \pm 0.01)\times10^{12}\,M_\odot$. Extending the integration to larger radii yields $(3.91 \pm 0.03)\times10^{12}\,M_\odot$ within 475~kpc and $(4.2 \pm 0.1)\times10^{12}\,M_\odot$ within 620~kpc, where the outermost apertures increasingly include a contribution from diffuse ICL.

For the $R_1$ and $R_{\mathrm{edge}}$ masses, we first take into account the different estimates of the stellar masses within the error range of $R_1$ and $R_{\mathrm{edge}}$. In addition, to account for the differences in M/L, we also assume a stellar mass uncertainty of $\delta_{\rm mass}=0.19$ dex, following the value derived by \citet{trujillo2020} for the E0--S0+ subsample, which is representative of massive early-type systems such as IC~1101. The stellar mass density profiles are scaled up and down accordingly, and the size indicators are recomputed in each realization. The final uncertainties are obtained from the dispersion of these realizations, effectively combining the effects of background fluctuations and stellar mass normalization.

\section{Discussion}\label{sec:discussion}

Although Abell~2029, the cluster where IC~1101 resides, is traditionally regarded as one of the most dynamically relaxed clusters in the local Universe \citep[e.g.,][]{dressler1979,buote1996}, recent ultra-deep X-ray observations have revealed that its intracluster medium (ICM) still preserves clear signatures of past dynamical activity. Very deep \textit{Chandra} data show the presence of one of the longest and most continuous sloshing spirals known, extending to nearly $\sim600$~kpc from the cluster core, together with additional substructures associated with a past off-axis merger and the subsequent settling of the system \citep{watson2026}. Earlier X-ray studies had already identified large-scale spiral-shaped residuals extending several hundred kiloparsecs from the cluster centre, interpreted as the remnant of an interaction with a subcluster or group 2-3 gigayears ago \citep{paterno2013}. In addition, IC~1101 is known to host a C-shaped radio jet in its core, indicative of gas sloshing in the central regions \citep{taylor1994}. Taken together, these results indicate that Abell~2029, while appearing relatively relaxed at first glance, reveals a more complex dynamical state when examined in detail. In particular, deeper analyses in both the X-ray and optical regimes uncover signatures, such as extended spiral-like structures and disturbed radio morphology, that suggest the system is still in the process of reaching full dynamical equilibrium.

In this context, the complex network of LSB structures unveiled around IC~1101 after the careful correction of scattered light from both foreground stars and the galaxy itself provides a natural stellar counterpart to the dynamical complexity observed in the hot gas. As illustrated in Fig.~\ref{fig:full_color_image}, several diffuse stellar features are detected at large radii, some displaying morphologies suggestive of accretion-driven processes, such as elongated plumes, diffuse overdensities, and asymmetric envelopes (e.g. features labelled A--H). Notably, the broad ``splash'' of cooler gas identified in the deep \textit{Chandra} observations of Abell~2029 is found to spatially coincide with the regions where we detect prominent LSB features, in particular those labelled H and G. A direct visual comparison between the optical image and the residual X-ray emission map is presented in Appendix~\ref{appendix:xray_comparison}, where the X-ray contours are overlaid on the ultra-deep optical data. While a strict one-to-one spatial correspondence is not expected, the comparable spatial extent and coherent morphology of the diffuse stellar structures and the X-ray spiral residual suggest that both components may trace different phases of the same large-scale dynamical disturbance in the cluster core. A more detailed description of this comparison is provided in Appendix~\ref{appendix:xray_comparison}.

As an additional check, we also examined whether these diffuse features could be associated with foreground Galactic cirrus rather than with the IC~1101/Abell~2029 system. To this end, we inspected the $g-r$ colour map of the faint features in the field and compared the colours and spatial distribution of the filamentary structures with the typical optical colours of Galactic cirrus, following an approach similar to that of \citet{roman2020}. We also compared their morphology with the Galactic extinction maps of \citet{schlafly2011}. The observed colours and the lack of a spatial correspondence with the extinction structures do not support a Galactic-cirrus origin for the features labelled A--H. Although a detailed stellar-population analysis of each individual feature is beyond the scope of this paper, these checks support the interpretation that the detected LSB structures are associated with the IC~1101/Abell~2029 system.

The extreme structural extent of IC~1101 is naturally reflected in its location within the mass-size relation displayed in Fig.~\ref{fig:size_comparison}. Both $R_1$ and $R_{\mathrm{edge}}$ place the galaxy well within the upper envelope of the size--mass relations defined by early-type galaxies in the observational samples of \citet{trujillo2020} and \citet{chamba2022}. The stellar mass surface density measured at $R_{\mathrm{edge}} = 260$~kpc is fully consistent with values reported for galaxies of comparable morphological type, supporting the physical interpretation of this radius as a meaningful boundary of the main stellar body of the BCG.

The effective radius of IC~1101, $R_{\mathrm{e}} = 73 \pm 2$~kpc, is fully consistent with the upper envelope of the stellar mass--size relation shown in Fig.~\ref{fig:size_comparison}. This value is larger than some of the effective radii reported in earlier works \citep[e.g.,][]{fisher1995}, who derive a mean effective radius of $R_{\mathrm{e}} \sim 42 \pm 8$~kpc, although broadly compatible once differences in depth, profile fitting, and treatment of extended light are taken into account. The ultra-deep imaging presented here enables a more complete recovery of the LSB envelope, while the careful PSF treatment ensures that the size measurement is not biased by scattered light contamination. Importantly, regardless of the specific size estimator adopted ($R_{\mathrm{e}}$, $R_1$, or $R_{\mathrm{edge}}$), IC~1101 consistently occupies the extreme upper-right region of the mass-size relations, highlighting its exceptional spatial extent compared to the general population of early-type galaxies. This behaviour is further illustrated in Fig.~\ref{fig:r1vsredge}, where IC~1101 follows the same $R_{\mathrm{edge}}$--$R_1$ relation defined by the sample of \citet{chamba2022}, extending it toward the largest radii currently probed.

The stellar mass profile of IC~1101 confirms that its extreme spatial extent is accompanied by an exceptionally large stellar mass budget. The progressive increase of enclosed mass from $R_{\mathrm{edge}} = 260 \pm 38$~kpc, where the stellar mass amounts to $(3.36 \pm 0.01)\times10^{12}\,M_\odot$, to $R=$ 620~kpc, enclosing $(4.20 \pm 0.1)\times10^{12}\,M_\odot$, reflects the growing contribution of diffuse stellar components, transitioning from the main body of the BCG to an extended envelope and ICL. The increasing uncertainties at large radii further highlight the intrinsically LSB and complex nature of these outer regions, which are shaped by prolonged accretion and incomplete dynamical relaxation.

The stellar mass estimates derived in Sect.~\ref{sec:size_comparison} are broadly consistent with previous observational constraints. \citet{uson1991} measured a total luminosity of $1.02 \times 10^{12}\,L_\odot$ for the IC~1101 system (BCG+ICL), which corresponds to a stellar mass of $(3.0 \pm 0.3)\times10^{12}\,M_\odot$ within a radius of $\sim 600$~kpc when converted using our adopted mass-to-light prescriptions (Appendix~\ref{appendix:mass_conversion}). In addition, \citet{dullo2017} reported a stellar mass of $M_\star \sim 1.1 \times 10^{12}\,M_\odot$ for the spheroidal component alone, corresponding to approximately 25\% of the total stellar mass. Extrapolating from this fraction yields a total stellar mass of $\sim 4.4 \times 10^{12}\,M_\odot$, in agreement with our measurements. The convergence of these independent estimates, despite substantial differences in methodology, data quality, and radial coverage, provides robust evidence that IC~1101 is among the most massive and spatially extended stellar systems known.

\section{Conclusions}\label{sec: conclusions}

In this study, we performed an ultra-deep imaging and photometric analysis of IC 1101. Our goal was to characterise the extension of this massive system. Based on sharp structural changes in its radial profiles (surface brightness, surface mass density, colour, ellipticity, etc), this study provides a measurement of the galaxy’s edge. Our main findings can be summarized as follows:

\begin{itemize}
    \item We modelled the extended PSF of the INT/WFC using a hybrid approach combining unsaturated and saturated stars, enabling accurate subtraction of scattered light and the detection of structures fainter than $\mu_r \sim 30$ mag arcsec$^{-2}$.

    \item By applying wavelet-based scattered light correction and model-based deconvolution, we obtained corrected images from which we extracted radial surface brightness, colour, and stellar mass surface density profiles using both free ellipses and a wedge-shaped region aligned with the galaxy's major axis.

    \item The analysis of the radial profiles indicates that the boundary of the main stellar body of the BCG is located at an edge radius of $R_{\mathrm{edge}} = 260 \pm 38$~kpc. In addition, an inner transition is identified at $\sim146$~kpc, associated with a flattening of the colour gradient, which suggests a regime of increased mixing of stellar populations. Beyond the edge radius, the stellar envelope extends to substantially larger scales, reaching $\sim 475$~kpc, where the contribution from diffuse ICL becomes increasingly important. The edge radius measured for IC~1101 remains among the largest values reported to date for an individual galaxy, corresponding to a projected diameter of $\sim 520$~kpc. The stellar mass enclosed within $R_{\mathrm{edge}}$ amounts to $(3.36 \pm 0.01)\times10^{12},M_\odot$, while the mass enclosed within the inner transition at $\sim150$~kpc is $(2.662 \pm 0.004)\times10^{12},M_\odot$. The mass within $R_1 = 362 \pm 38$~kpc is $(3.69 \pm 0.01)\times10^{12},M_\odot$. Extending the integration to larger radii yields $(3.91 \pm 0.03)\times10^{12},M_\odot$ within 475~kpc and $(4.2 \pm 0.1)\times10^{12},M_\odot$ within 620~kpc, where the outermost apertures increasingly include a contribution from diffuse ICL and LSB structures. Adopting the stellar mass enclosed within the radius at which the $g$-band surface brightness reaches $\mu_g = 29$~mag~arcsec$^{-2}$, we obtain a total stellar mass of $(4.2 \pm 0.1)\times10^{12},M_\odot$, placing IC~1101 at the upper extreme of the mass--size relation.

    \item The spatial properties of the southwestern LSB structure are consistent with the X-ray spiral residuals reported by \citet{paterno2013} and \citet{watson2026}, which have been interpreted as the outcome of an off-axis merger with a subcluster approximately $2$--$3$~Gyr ago.
\end{itemize}

\begin{acknowledgements}

The authors thank the anonymous referee for their thoughtful and constructive comments, which helped to improve the quality and clarity of this manuscript. The authors gratefully acknowledge Juan Uson for his valuable contributions and insightful comments on this work. This work has made use of data from the European Space Agency (ESA) mission {\it Gaia} (\url{https://www.cosmos.esa.int/gaia}), processed by the {\it Gaia} Data Processing and Analysis Consortium (DPAC, \url{https://www.cosmos.esa.int/web/gaia/dpac/consortium}). Funding for the DPAC has been provided by national institutions, in particular the institutions participating in the {\it Gaia} Multilateral Agreement. This project is possible thanks to financial support from the Spanish Ministry of Science and Innovation (MICINN) to the coBEARD project (PID2021-128131NB-I00). 

JMA acknowledges the support of the Viera y Clavijo Senior program funded by ACIISI and ULL and the support of the Agencia Estatal de Investigación del Ministerio de Ciencia e Innovación (MCIN/AEI/10.13039/501100011033) under grant nos. PID2021-128131NB-I00 and CNS2022-135482 and the European Regional Development Fund (ERDF) ‘A way of making Europe’ and the ‘NextGenerationEU/PRTR’. 

AdLC  acknowledges financial support from the Spanish Ministry of Science and Innovation (MICINN) through RYC2022-035838-I and PID2021-128131NB-I00 (CoBEARD project). 

IT acknowledges support from the State Research Agency (AEI-MCINN) of the Spanish Ministry of Science and Innovation under the grant PID2022-140869NB-I00 and IAC project P/302302, financed by the Ministry of Science and Innovation, through the State Budget and by the Canary Islands Department of Economy, Knowledge, and Employment, through the Regional Budget of the Autonomous Community. This research also acknowledge support from the European Union through the following grants: "UNDARK" and "Excellence in Galaxies - Twinning the IAC" of the EU Horizon Europe Widening Actions  programmes (project numbers 101159929 and 101158446). Funding for this work/research was provided by the European Union (MSCA EDUCADO, GA 101119830).

FB and SR acknowledge the support of the grants PID2023-150393NB-I00 and CNS2024-154572 from the Spanish Ministry of Science, Innovation, and Universities. Financial support of the Department of Education, Junta de Castilla y Le\'{o}n, and FEDER Funds is gratefully acknowledged (Reference: CLU-2023-1-05). 

MM acknowledges support from RYC2022-036949-I financed by the MICIU/AEI/10.13039/501100011033 and by ESF+. MM and ZS acknowledge support from grant CNS2024-154592 financed by MICIU/AEI/10.13039/501100011033, and program Unidad de Excelencia Mar\'{i}a de Maeztu CEX2020-001058-M, financed by MCIN/AEI/10.13039/501100011033, and by the MaX-CSIC Excellence Award MaX4-SOMMA-ICE.

SGA, IRC and GG acknowledge support from grant PID2022-140869NB-I00 from the Spanish Ministry of Science and Innovation. 

RIS acknowledges financial support from the Spanish Ministry of Science and Innovation through the project PID2022-138896NAC54.

AAR acknowledges funding from the Agencia Estatal de Investigación del Ministerio de Ciencia, Innovaci\'on y Universidades (MCIU/AEI) under grant ``Polarimetric Inference of Magnetic Fields'' and the European Regional Development Fund (ERDF) with reference PID2022-136563NB-I00/10.13039/501100011033.

JR acknowledges financial support from the Plan Propio de Investigación 2025 submodalidad 2.3 of the University of Córdoba.

\end{acknowledgements}

\bibliography{references.bib}

\begin{appendix}

\renewcommand{\thefigure}{A.\arabic{figure}}
\setcounter{figure}{0} 
\begin{figure*}[t!]
    \centering
    \includegraphics[width =0.9\linewidth]{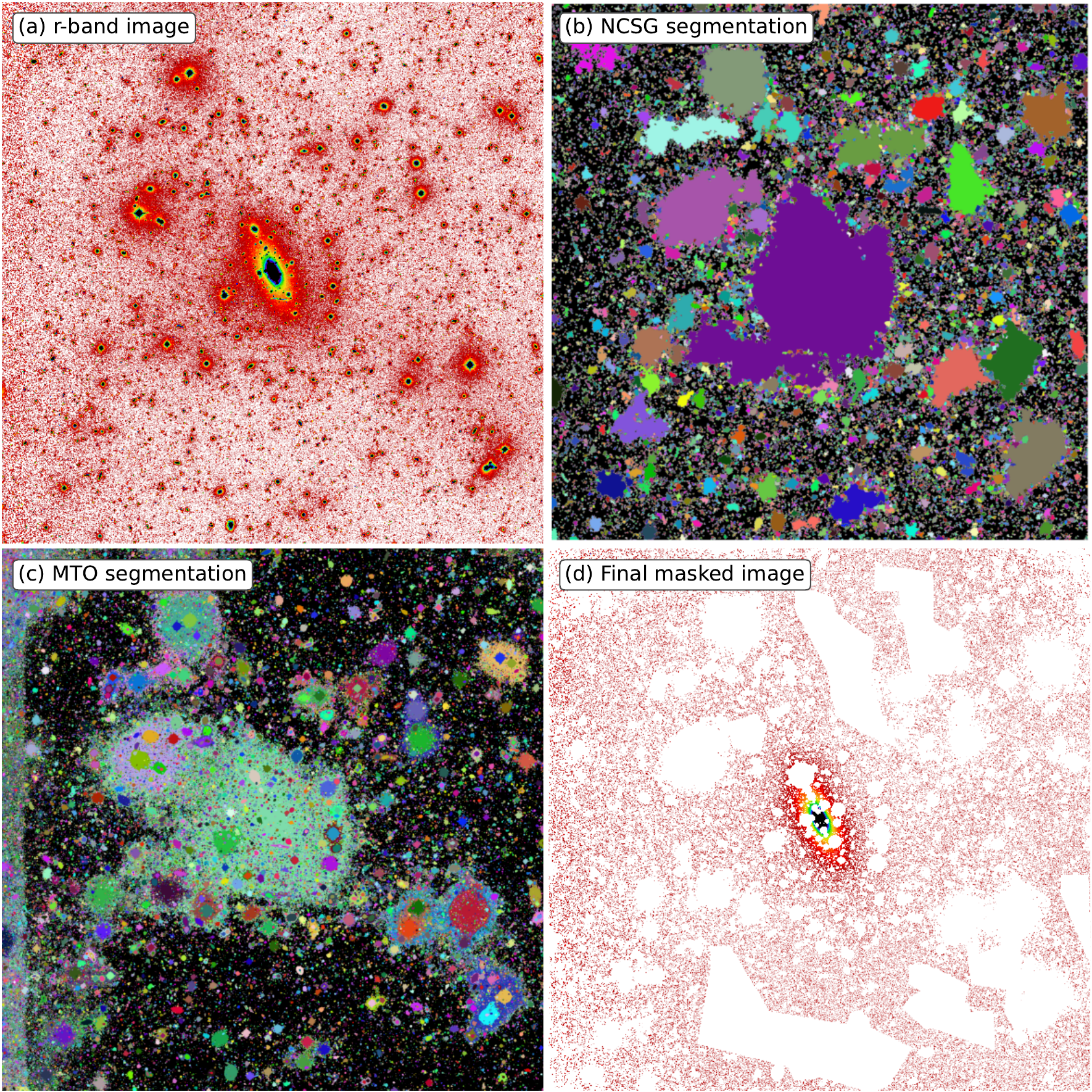}
    \caption{Comparison of the masking procedure applied to the $r$-band image of IC~1101. Panel (a) shows the original $r$-band image using the same intensity scaling as in the final masked image. Panel (b) shows the segmentation map produced with \texttt{NoiseChisel} and \texttt{Segment} from the GNU Astronomy Utilities suite (NCSG). Panel (c) shows the segmentation map generated using the \texttt{MTO} algorithm. Panel (d) shows the final masked $r$-band image, where the masked regions are shown in white. This final mask combines the automatic masks with additional manually defined masks.}

    \label{fig:segmentation_comparison}
\end{figure*}

\section{Segmentation strategies for masking contaminant sources}
\label{appendix:segmentation}

In order to mask foreground and background sources in the field surrounding IC~1101, we employed two independent segmentation strategies: the combination of \texttt{NoiseChisel} and \texttt{Segment} \citep{akhlaghi2015, akhlaghi2019} from the GNU Astronomy Utilities (GNUastro) suite, and the object detection algorithm known as \texttt{MTO} (Max-Tree Objects; \citealt{teeninga2015,faezi2024}). These methods were selected for their complementary behaviour and for their different roles in the analysis. The GNUastro-based segmentation was mainly used in the technical stages of the reduction and image treatment, in particular for the construction of the extended PSF and the modelling and subtraction of scattered light from bright stars. In contrast, the \texttt{MTO}-based segmentation was adopted as the primary masking strategy for the scientific analysis of IC~1101, including the extraction of surface-brightness, colour, and stellar-mass-density profiles.

In our implementation, the following parameters were adopted to optimise the detection in the low-surface-brightness regime: \texttt{NoiseChisel (-{}-tilesize=20,20 -{}-interpnumngb=5 -{}-dthresh=0.05 -{}-snminarea=2)} and \texttt{Segment (-{}-tilesize=10,10 -{}-interpnumngb=1 -{}-gthresh=-10 -{}-objbordersn=1 -{}-minnumfalse=1)}. These segmentation maps were used to support the PSF construction and scattered-light correction, where the main requirement is to obtain a robust and reproducible identification of contaminating sources across the full image.

For the analysis phase, particularly for the extraction of surface-brightness, colour, and stellar-mass-density profiles, we required a segmentation capable of reliably deblending overlapping sources and isolating compact contaminants projected onto the extended galaxy emission. For this purpose, we adopted the \texttt{MTO} algorithm, which is well suited to separating superimposed sources in crowded and LSB environments, such as the outskirts of massive galaxies like IC~1101. We found that the default \texttt{MTO} configuration already provides an adequate segmentation for this task, with the exception of setting \texttt{-{}move\_factor=0}.

The final mask used for the scientific measurements was constructed by combining the \texttt{MTO} masks obtained independently in the $g$- and $r$-band images. This ensures that sources detected in either filter are consistently excluded from the analysis and reduces possible residual contamination associated with colour-dependent detection effects. In addition, we included a set of manually defined masks in regions where visual inspection indicated that the automatic segmentation could be incomplete, for instance around residual stellar haloes, extended contaminants, saturated sources, or regions affected by complex overlapping structures. These manual additions were only used to refine the final mask in specific problematic areas and to ensure a conservative treatment of contaminating sources.

Fig.~\ref{fig:segmentation_comparison} presents a visual comparison of the masking procedure applied to the $r$-band image of IC~1101. The figure shows the original image, the segmentation map obtained with \texttt{NoiseChisel} and \texttt{Segment}, the segmentation map produced with \texttt{MTO}, and the final masked image used in the scientific analysis. As highlighted by \citet{haigh2021}, segmentation algorithms, including \texttt{NoiseChisel} and \texttt{MTO}, show complementary behaviour, particularly when dealing with nested or overlapping sources. Their relative performance can depend on the nature of the structures being deblended and on the specific parameter configurations adopted. For this reason, the combination of GNUastro-based segmentation for the technical image-treatment steps and \texttt{MTO}-based masks for the final scientific measurements provides a robust and flexible masking strategy for the analysis of IC~1101.

\section{Catalogue creation and star selection} 
\label{appendix:catalogs_PSF}

\renewcommand{\thefigure}{B.\arabic{figure}}
\setcounter{figure}{0} 
\begin{figure}[t]
    \centering
    \includegraphics[width =\columnwidth]{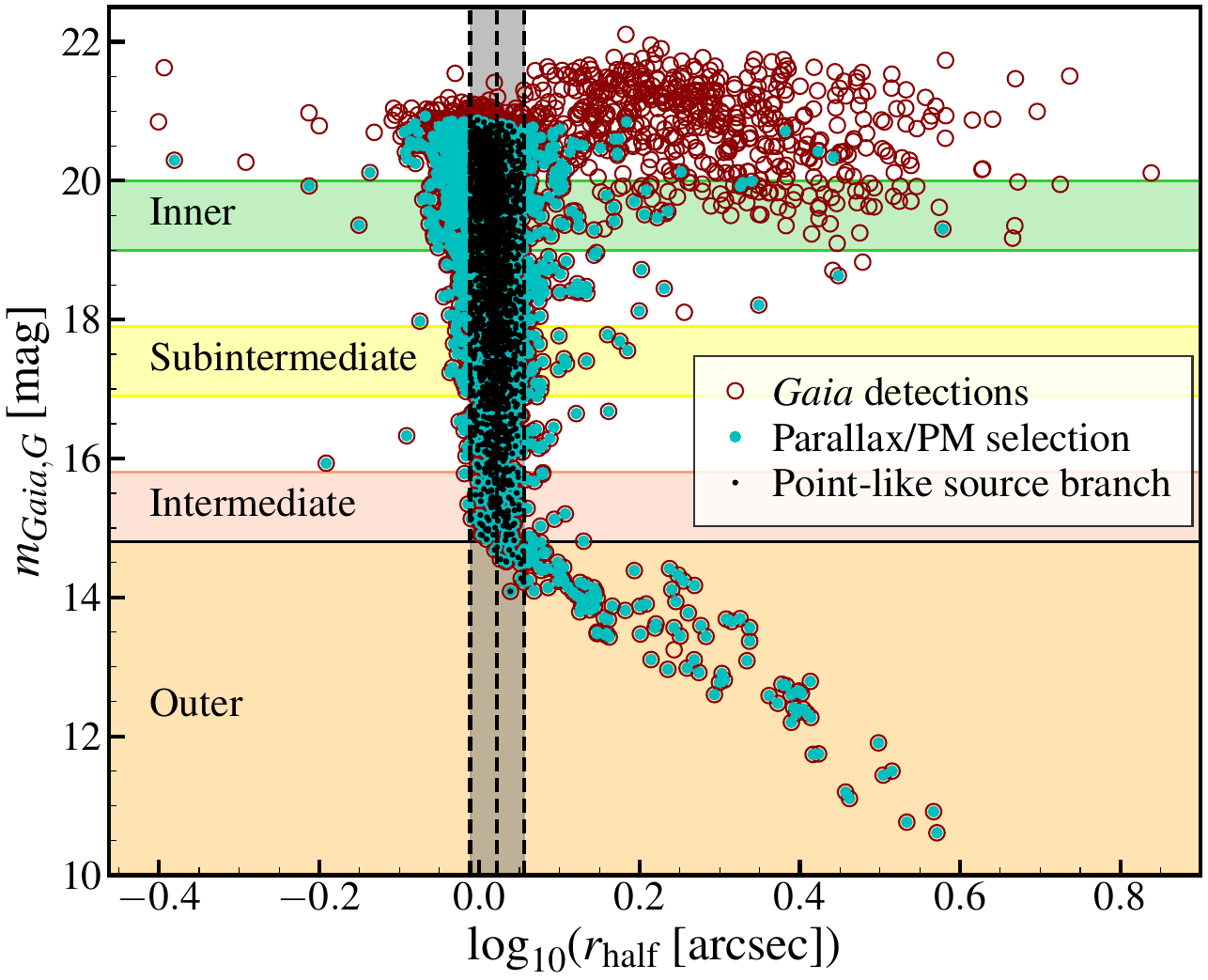}
    \caption{ Point-like source selection diagram showing the \textit{Gaia} $G$-band magnitude ($m_{Gaia,G}$) as a function of the logarithm of the half-sum-radius ($r_{\mathrm{half}}$) in arcsec for all detected sources. Red open circles indicate all \textit{Gaia} cross-matched detections, while cyan filled circles represent sources that passed the parallax or proper motion significance filter (see text for details). Black dots highlight the sources identified as part of the point-source branch, selected via a $3\sigma$ sigma-clipping. Vertical dashed black lines denote the range of $r_{\mathrm{half}}$ values used to select these stars. The horizontal coloured bands correspond to different magnitude intervals employed for the characterization of specific radial regions of the PSF: inner (green), subintermediate (yellow), intermediate (salmon), and outer (orange).} 
    \label{fig:HSM_Mag}
\end{figure}

Using the detections obtained during the masking procedure described in Sect.~\ref{section:image_masking} and Appendix~\ref{appendix:segmentation}, we constructed a catalogue of sources based on the segmentation maps produced by \texttt{NoiseChisel} and \texttt{Segment}. This catalogue was used to identify suitable stars (point-like sources) for the construction of the large-scale PSF and to generate the corresponding photometric and positional information.

To refine the stellar sample, the GNUastro-based catalogue was cross-matched with the \textit{Gaia} Data Release 3 (DR3) catalogue (\citealt{gaia2016}; \citealt{gaia2023}). We tested several matching radii and found that a value of 5$\arcsec$ provided a good compromise: it is large enough to recover the most saturated and brightest stars, essential for constraining the outer wings of the PSF, while still minimizing false associations.

After cross-matching, we retain only those sources for which the parallax or proper motions (in right ascension and declination) exceed three times their associated uncertainties. In other words, we keep only bona fide stellar objects with well-constrained astrometric parameters. To identify the final stellar sample, we used the half-sum radius ($r_{\mathrm{half}}$) provided by \texttt{GNUastro} as a proxy for the size of the sources. This parameter, derived from the area enclosing half the total flux of each labelled region, offers a robust proxy for the size for point-like sources. In Fig.~\ref{fig:HSM_Mag}, we plot the \textit{Gaia} $G$-band magnitude ($m_{Gaia,G}$) against $r_{\mathrm{half}}$ and construct a diagnostic diagram in which point-like sources naturally define a narrow vertical band. The advantage of using \textit{Gaia}’s $G$-band, as discussed by \citet{sedighi2024}, is that its photometry is less affected by saturation and provides a uniform, high-quality reference for sources spanning a wide range of magnitudes.

The following procedure was applied to the $r_{\mathrm{half}}$ distribution shown in Fig.~\ref{fig:HSM_Mag}. We first applied a $2\sigma$ clipping to remove outliers and computed a mean $r_{\mathrm{half}}$ value from the remaining sources. The final stellar sample was then defined as those objects lying within $3\sigma$ of this mean, ensuring a homogeneous selection of stars with consistent structural properties.

\renewcommand{\thefigure}{B.\arabic{figure}}
\setcounter{figure}{0} 

\section{PSF joining regions and external extrapolation} 
\label{appendix:regions_PSF}

The specific transition points between consecutive regions were determined based on the highest S/N within each radial interval. From the innermost to the outermost regions, the radii with the highest S/N ratios are 3.2\arcsec\ (Inner–Subintermediate), 5.8\arcsec\ (Subintermediate–Intermediate), 9.8\arcsec\ (Intermediate–Outer), and 115.5\arcsec\ (Outer–Outer Ext).

A PSF should be at least twice the size of the object studied \citep{sandin2014, sandin2015}; therefore, we extended our constructed PSFs to cover the full radial range required to match the size of IC~1101. In both bands, the PSF wings were modelled as a power-law decline of the form $I(r) \propto r^{\alpha}$. The extrapolation followed the rationale of \citet{sedighi2024}, who emphasize the need for a fitting window that provides a compromise between having high S/N in the outer PSF profile and avoiding over-subtraction of the background that may affect the faintest regions. In our case, the extrapolation was performed over the radial range between approximately $300''$ and $460''$ in the $g$ band, and between $\sim400''$ and $500''$ in the $r$ band. These intervals were selected to correspond to regions where the azimuthally averaged PSF profile still remains significantly detected above the background, with typical S/N ratios of $\sim6$ in the $g$ band and $\sim4$ in the $r$ band. Both profiles were then extended to a final PSF radius of $\sim37.4$~arcmin ($2244''$), ensuring complete coverage of the galaxy and its surroundings. The best-fit slopes obtained are $\alpha_g = -2.5\pm0.1$ and $\alpha_r = -2.7\pm0.3$ for the $g$- and $r$-bands, respectively. The quoted uncertainties were estimated by slightly varying the radial fitting window by a few pixels toward smaller and larger radii, repeating the power-law fit in each case, and taking the resulting variation in the slope as an estimate of the systematic uncertainty associated with the choice of fitting range. These values are in good agreement with previous measurements of extended PSF profiles in deep imaging. For instance, \citet{infante2020} report power-law slopes of $\sim -2.5$ for SDSS PSFs, while \citet{sedighi2024} find slightly shallower slopes of $\sim -2.1$ using LBT data.

\section{Scattered light map}
\label{appendix:scatter light map}

In order to match the PSF model profile with those of the real stars in the image, a radial range (annulus) must be found where the profiles match. This annulus is defined using fixed surface-brightness thresholds applied to the PSF-scaled stellar profiles. We first build a two-dimensional lookup table that maps stellar magnitude and surface brightness to a characteristic radius, $R_{\textrm{char}}(m,\mu)$. The table is computed directly from the measured PSF radial profile: we integrate the PSF to obtain its total flux, derive the multiplicative scale factor required to reproduce a given stellar magnitude, and then determine the radius at which the scaled PSF reaches a chosen surface-brightness level. This provides a fast and self-consistent estimate of $R_{\textrm{char}}(m,\mu)$ without repeatedly rescaling and re-interpolating the PSF for each star. The resulting magnitude-dependent radii for $\mu=21$ and $24$~mag~arcsec$^{-2}$ are shown in Fig.~\ref{appendix:sat_radius}, where the shaded region illustrates the annular area adopted for the PSF normalization.

\renewcommand{\thefigure}{D.\arabic{figure}}
\setcounter{figure}{0} 
\begin{figure}[t]
    \centering
    \includegraphics[width =\columnwidth]{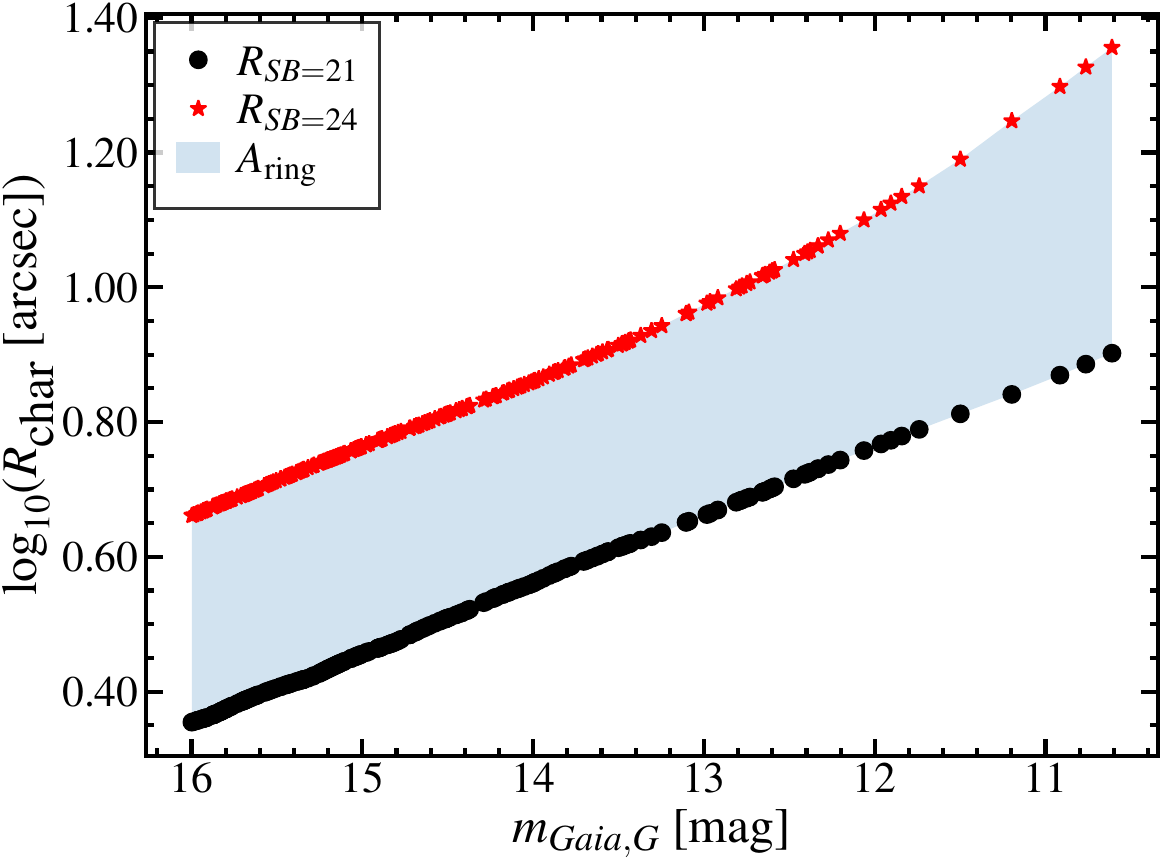}
    \caption{Magnitude-dependent definition of the PSF normalization annulus. The black circles and red stars show, respectively, the characteristic radii $R_{\mathrm{SB}=21}$ and $R_{\mathrm{SB}=24}$ at which the PSF-scaled stellar profiles reach surface-brightness levels of $\mu=21$ and $24$~mag~arcsec$^{-2}$. The vertical axis shows $\log_{10}(R_{\mathrm{char}})$ in arcsec, while the horizontal axis indicates the \textit{Gaia} $G$-band magnitude. The shaded region corresponds to the annular area $A_{\mathrm{ring}}$ used to compute the multiplicative scaling factor between each star and the reference PSF. The radii are derived from the precomputed magnitude--surface-brightness--radius lookup table built from the PSF model, ensuring that the normalization is performed in a uniform surface-brightness regime that is safely outside the saturated core and above the noise-dominated outskirts. The annulus therefore shifts systematically outward for brighter stars and inward for fainter ones, naturally adapting to stellar brightness while maintaining consistent normalization conditions across the full sample.}
    \label{appendix:sat_radius}
\end{figure}

\renewcommand{\thefigure}{D.\arabic{figure}}
\setcounter{figure}{1} 
\begin{figure*}[t!]
    \centering
    \includegraphics[width =1.8\columnwidth]{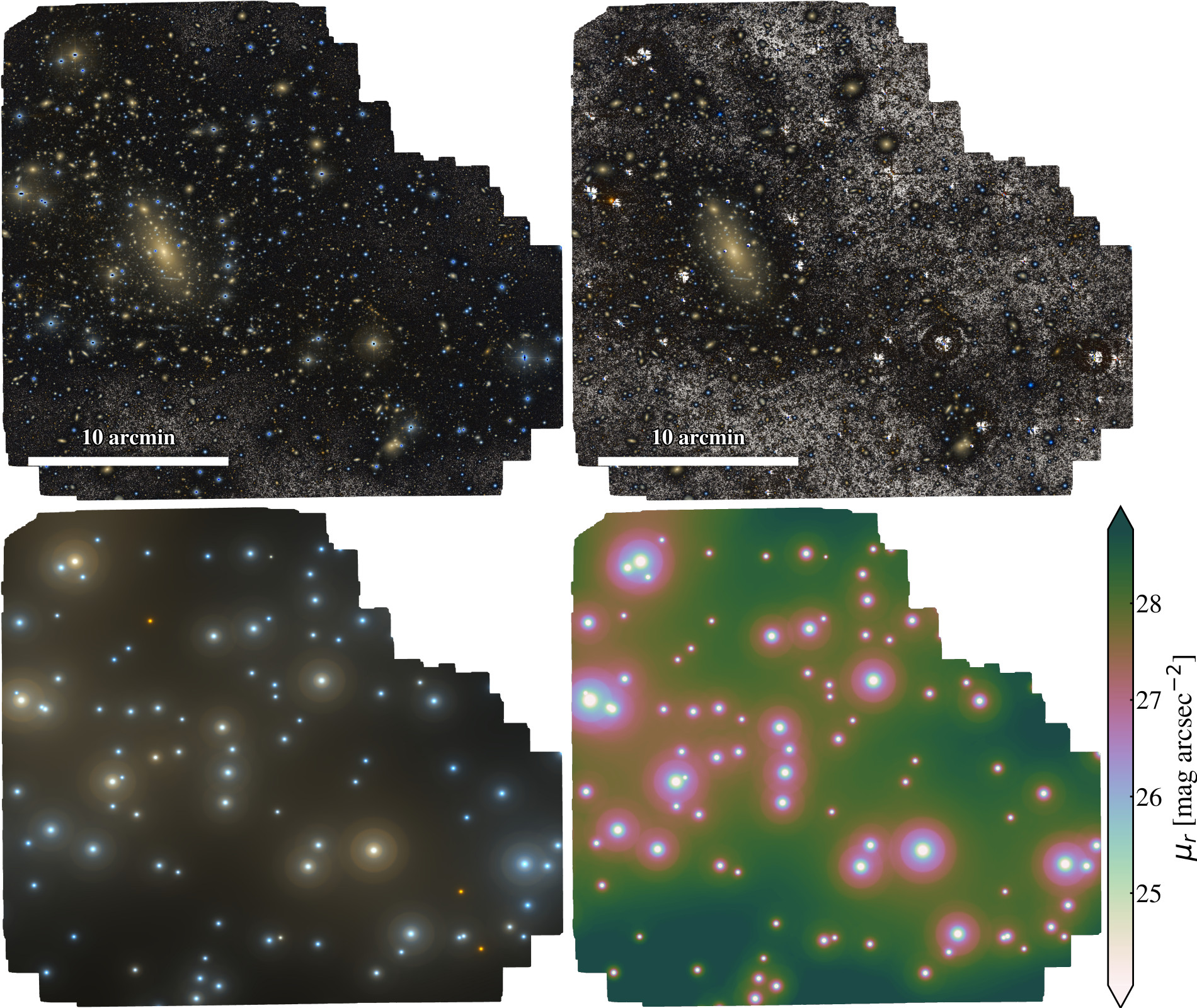}
    \caption{
    Two--by--two comparison of the scattered--light field around IC\,1101.
    Top panels show the original RGB composite (left) and the same field after foreground--star subtraction (right).
    Bottom panels display the RGB composite of the scattered--light field (left) and the scattered--light map in the $r$-band produced by \texttt{MAHDI} (right), in surface--brightness units. The scattered--light map reveals the extended halos of bright stars and their contamination around the galaxy.
    }
   \label{fig:comparison_scatter_map}
\end{figure*}

For the flux normalization we adopt $\mu=21$ and $24$~mag~arcsec$^{-2}$, which bracket a conservative high-S/N regime while remaining safely outside the saturated cores. In particular, the brightest stars in our $g$-band data (the deepest of the two filters) reach central surface brightness values of $\mu_{0,g}=17.7\pm0.5$~mag~arcsec$^{-2}$ on average, supporting the use of $\mu=21$--24~mag~arcsec$^{-2}$ as a robust normalization interval. For each star, the inner and outer radii of the matching annulus are then set to $R_{\rm SB=21}=R_{\textrm{char}}(m,\mu=21)$ and $R_{\rm SB=24}=R_{\textrm{char}}(m,\mu=24)$, respectively, and the PSF scaling factor is computed from the flux measured within this annulus. As illustrated in Fig.~\ref{appendix:sat_radius}, this procedure naturally shifts the normalization region outward for brighter stars and inward for fainter ones, while enforcing a uniform surface-brightness domain for the PSF matching across the full stellar sample.

In our analysis, we constructed a model of the scattered--light field in the $g$- and $r$-bands using the methodology described in Section~\ref{sec:subtract_stars}. Each star was modelled using the extended PSF, and all stellar models were subsequently combined to produce a smooth, empirical map of the scattered--light contribution across the field. Fig.~\ref{fig:comparison_scatter_map} presents a visual comparison in which the top panels show the original RGB composite image of the galaxy (left) and the same field after the subtraction of foreground stars (right). The latter clearly reveals a wealth of LSB structures that become apparent only after this treatment. The bottom panels display the RGB composite of the scattered--light field (left) and the corresponding scattered--light map in the $r$-band (right), expressed in surface--brightness units. The scattered--light map highlights the widespread contamination induced by extended stellar haloes, particularly in the vicinity of the galaxy. In this region, the scattered--light screen reaches surface--brightness levels of $\mu_r \sim 27$--$28$~mag~arcsec$^{-2}$, making the subtraction of stars essential for a reliable analysis of the galaxy's faint stellar envelope.

\section{Wavelet framework}\label{sec:wave_framework}

In the context of image deconvolution, we incorporate wavelet-based regularization \citep{starck2010,carrillo2012} as a means to suppress noise amplification and preserve meaningful astrophysical structures across different spatial scales. This is achieved by penalizing the contribution of wavelet coefficients during the optimization process, effectively discouraging the introduction of non-physical features into the deconvolved image. Specifically, the algorithm operates by minimizing a composite loss function that combines a data fidelity term, which ensures consistency with the observed image, and one or more regularization terms. The optimization is carried out iteratively using a first-order method, specifically the AdamW optimizer \citep{loshchilov2017}, implemented within the PyTorch framework \citep{pythorchcite}. The model parameters are updated in the direction that minimizes the overall loss. This approach provides a flexible and stable framework for adaptive deconvolution in the presence of extended PSFs and low S/N conditions.

For the wavelet-based losses, we employ the Isotropic Undecimated Wavelet Transform (IUWT; \citealt{starck2002}), chosen for its isotropy and shift-invariance properties, which make it particularly suitable for astrophysical applications. The IUWT decomposes the image into a set of scale-dependent bands without decimation, preserving the original image size at each level. The corresponding regularization term is defined as:
\begin{equation}
\mathcal{R}_{\mathrm{IUWT}} = \sum_{j=1}^{J} \alpha_j \left\| w_j(x, y) \right\|_1,
\end{equation}

where \(w_j(x, y)\) is the coefficient map at scale \(j\), and \(\alpha_j\) is a scale-dependent weight proportional to the expected noise level at that scale, estimated via Monte Carlo sampling.

These wavelet-based penalties are integrated into the total loss function minimized during the iterative deconvolution process:

\begin{equation}
\mathcal{L}_{\mathrm{total}} = \mathcal{L}_{\mathrm{MSE}} + \lambda_{\mathrm{grad}} \mathcal{R}_{\mathrm{grad}} + \lambda_{\mathrm{obj}} \mathcal{R}_{\mathrm{obj}} + \lambda_{\mathrm{wave}} \mathcal{R}_{\mathrm{wave}}\\ 
\end{equation}
\begin{equation*}
    + \lambda_{\mathrm{IUWT}} \mathcal{R}_{\mathrm{IUWT}} + \lambda_{1} \mathcal{R}_{L_1}.
\end{equation*}

Each term in the loss function serves a specific regularization purpose:

\begin{itemize}
    \item \(\mathcal{L}_{\mathrm{MSE}} = N^{-1} \sum (\mathrm{obs} - \mathrm{convolved})^2\): The mean squared error between the observed and model-predicted images, representing the main likelihood term.\\
    
    \item \(\mathcal{R}_{\mathrm{grad}} = \lambda_{\mathrm{grad}} \cdot \mathrm{mean}(\nabla I)^2\): Penalizes large spatial gradients to suppress noise and enforce smoothness.\\
    
    \item \(\mathcal{R}_{\mathrm{obj}} = \lambda_{\mathrm{obj}} \cdot \mathrm{mean}(I^2)\): Penalizes large pixel amplitudes to prevent unphysical brightness peaks.\\
    
    \item \(\mathcal{R}_{\mathrm{wave}} = \lambda_{\mathrm{wave}} \cdot \mathcal{L}_{\mathrm{wave}}(I)\): Penalizes high-frequency structures detected in the DWT to control small-scale noise.\\
    
    \item \(\mathcal{R}_{L_1} = \lambda_{1} \cdot \mathrm{mean}(|I|)\): Enforces sparsity in pixel intensities via L1 norm regularization.\\
    
    \item \(\mathcal{R}_{\mathrm{IUWT}} = \lambda_{\mathrm{IUWT}} \cdot \mathcal{L}_{\mathrm{IUWT}}(I)\): Penalizes structure at specific scales in the IUWT, helping to control diffuse, large-sc
ale artifacts.
\end{itemize}

\renewcommand{\thefigure}{F.\arabic{figure}}
\setcounter{figure}{0} 

\begin{figure}[t!]
    \centering
    \includegraphics[width=\columnwidth]{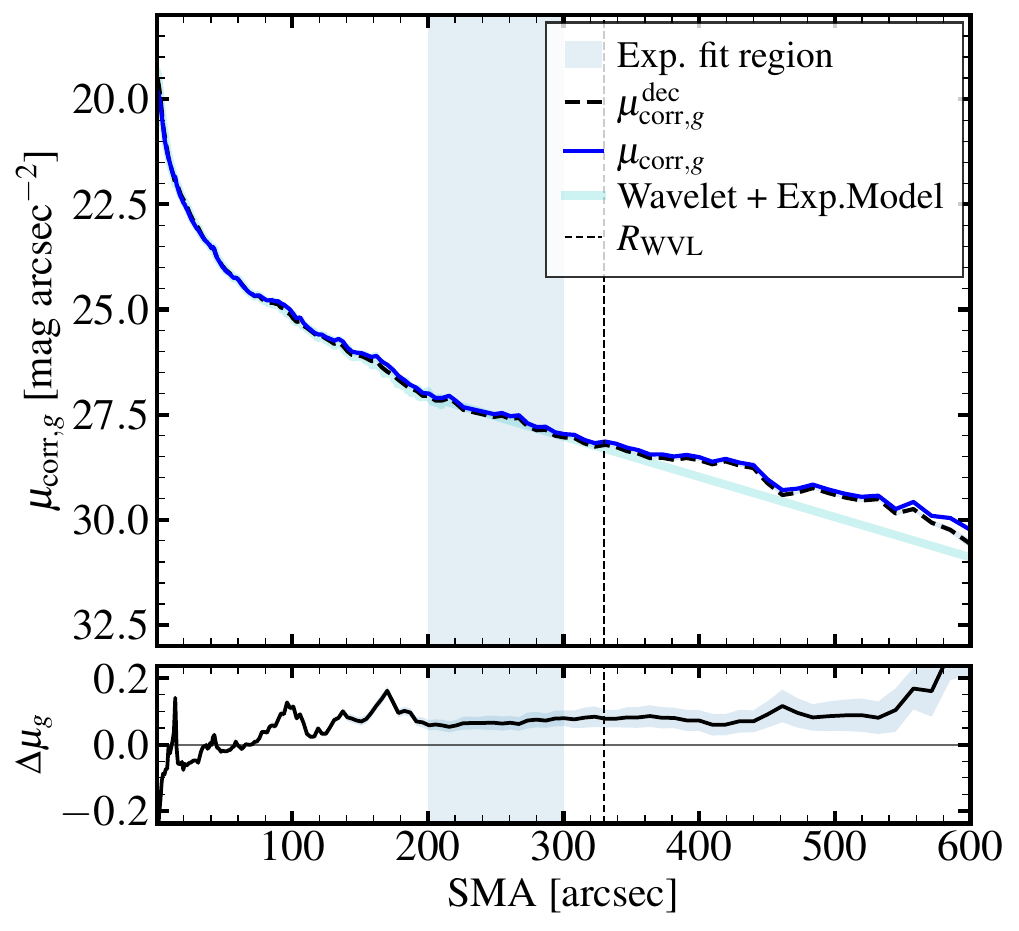}
    \caption{Surface brightness profiles of IC~1101 in the $g$-band before and after the wavelet-based PSF correction. The upper panel shows the corrected profile $\mu_{\mathrm{corr},g}$ (solid blue curve) and the deconvolved profile $\mu_{\mathrm{corr},g}^{\mathrm{dec}}$ (dashed black curve). The shaded region marks the radial range (200–300 arcsec) used to fit an exponential model, which is extrapolated to larger radii (cyan curve) to represent the expected symmetric decline in the absence of PSF scattering. The vertical dashed line indicates $R_{\mathrm{WVL}}$, beyond which the profile departs from the exponential trend due to the presence of asymmetric LSB structures. The lower panel shows the residual $\Delta\mu_g = \mu_{\mathrm{corr},g}^{\mathrm{dec}} - \mu_{\mathrm{corr},g}$, quantifying the magnitude difference between both profiles, with the shaded region indicating the associated uncertainties.}
    
    \label{fig:sb_deconvolution_comparison}
\end{figure}

For the results presented in this work, the regularization weights were set to $\lambda_{\mathrm{grad}} = 10^{-6}$, $\lambda_{\mathrm{obj}} = 10^{-2}$, $\lambda_{\mathrm{wave}} = 10^{-6}$, $\lambda_{1} = 10^{-6}$, and $\lambda_{\mathrm{IUWT}} = 10^{-2}$. These values were chosen empirically to balance noise suppression and structure preservation, and were kept fixed throughout the analysis. This combination of data fidelity and multiscale regularization ensures that the deconvolved images remain physically plausible, while effectively mitigating instrumental effects and residual scattered light. The regularization weights $\lambda$ are user-defined hyperparameters, and the total loss function is minimized iteratively using a first-order optimization scheme based on the \texttt{AdamW} optimizer, with optional mixed-precision acceleration. Each regularization term can be enabled or disabled independently, providing a flexible and modular framework for adaptive deconvolution of astronomical images. The deconvolution process follows the sequence:

\begin{enumerate}
    \item Input preparation: The observed image and the PSF are zero-padded symmetrically. Saturated regions (i.e., NaN pixels) can optionally be inpainted using locally scaled PSF cutouts centred at each saturated region.

    \item PSF normalization and Fourier transform: The PSF is normalized to unit flux and transformed into Fourier space. A low-pass circular mask is also generated in Fourier space based on a user-defined diffraction limit.

    \item Initialization: The initial image guess is taken as a clone of the input observation. This tensor is set to require gradients and serves as the optimization variable.

    \item Noise characterization for IUWT: If IUWT regularization is enabled, a random noise realization is passed through the IUWT transform to estimate the scale-dependent standard deviations \(\alpha_j\), which weight each wavelet level.

    \item Optimization loop: For each iteration:
    \begin{enumerate}
        \item The current image estimate is optionally clamped to non-negative values.
        \item The image is apodized with a Hanning window and Fourier-transformed.
        \item A Fourier-domain filter is applied (either a simple diffraction mask or the Löfdahl-Scharmer filter; \citealt{lofdahl1994}).
        \item The filtered image is convolved with the PSF (via inverse FFT) and compared to the observed image to compute the MSE loss.
        \item Spatial gradients and wavelet coefficients are computed for regularization terms.
        \item The total loss is assembled as the sum of all components.
        \item The loss is backpropagated and the optimizer updates the image.
    \end{enumerate}

    \item Final output: Once convergence is reached, the image is optionally clamped again and returned, along with the final Fourier-filtered estimate and the loss evolution history.
\end{enumerate}

The whole code structure can be found in a \texttt{GitHub} \protect\footnotemark \footnotetext{\url{https://github.com/aasensio/Wavelet_deconvolution}} repository.

\section{Model-based PSF correction of the surface-brightness profiles}
\label{appendix:sb_deconvolution}

To correct for the contribution of scattered light from the extended PSF wings of IC~1101 and nearby sources, we adopt a model-based approach that combines wavelet filtering with parametric profile fitting, following the methodology introduced by \citet{golini2025}. This procedure is designed to mitigate PSF-induced contamination while preserving the diffuse LSB structures detected in the galaxy outskirts.

As a first step, a wavelet-based filtering (based on Appendix \ref{sec:wave_framework}) is applied to the $g$- and $r$-band images to suppress small-scale fluctuations and enhance the large-scale galaxy emission without introducing high-frequency noise. From these wavelet-filtered images, we extract radial surface-brightness profiles along the semimajor axis. In both bands, the profiles exhibit a exponential decline over the radial range $200'' \lesssim R \lesssim 300''$. This behaviour is consistent with previous studies of IC~1101, which report the presence of an exponential stellar halo dominating at intermediate radii \citep{dullo2017}.

We therefore fit an exponential model to the deconvolved profiles within this radial interval. At larger radii, the profiles deviate from a simple exponential form and become shallower, reflecting the presence of asymmetric LSB structures already discussed. To avoid modelling these non-axisymmetric features, the exponential fit obtained between $200''$ and $300''$ is extrapolated outward and used as a reference description of the expected symmetric light distribution in the absence of PSF effects.

This exponential fit is used to construct a two-dimensional model using the global ellipticity and position angle of IC~1101. This model is combined with the wavelet-filtered inner region ($R < 200''$) and subsequently convolved with the extended PSF derived in Sect.~\ref{sec:INT_PSF}. The convolved model represents the contribution of PSF-scattered light associated with the galaxy itself and nearby extended sources. This component is subtracted from the original image, and the residuals are added back to the PSF-corrected model, yielding a final reconstruction that minimizes PSF contamination while retaining genuine diffuse structures.

Figure~\ref{fig:sb_deconvolution_comparison} shows that the overall impact of PSF scattering on the surface-brightness profile of IC~1101 is relatively low, in agreement with expectations from previous studies \citep{tal2011,montes2014}. This is further illustrated in the lower panel, where the residual $\Delta\mu_g$ remains typically within $\sim 0.05$--$0.1$ mag over most of the radial range, indicating that the intrinsic light distribution is already well recovered after the star-subtraction procedure. The deconvolved profile closely follows the observed one at all radii, with only small, localized deviations. In particular, around $\sim$100 and $\sim$160 arcsec, minor bumps in the observed profile are noticeably reduced after the wavelet-based deconvolution. These features are likely associated with residual scattered light from nearby sources which, despite being masked, still contaminate the profiles through the extended PSF wings.

Within the characteristic radius $R_{\mathrm{WVL}}$, marked by the vertical dashed line, the wavelet-based deconvolution is retained. At larger radii, where asymmetric LSB features dominate, the correction transitions to the hybrid model-based approach in order to preserve genuine diffuse structures in the galaxy outskirts. A detailed illustration and discussion of this hybrid strategy can be found in Figs.~2, 3, and 4 of \citet{golini2025}.

\section{Comparison with residual X-ray emission}
\label{appendix:xray_comparison}

In order to explore the possible connection between the diffuse optical structures identified around IC~1101 and the hot ICM, we compare our ultra-deep optical image with the residual X-ray emission map presented by \citet{watson2026}. 

The green contours overlaid in Fig.~\ref{fig:xray_overlay} correspond to the residual X-ray emission in the 0.7--7 keV band reported in that work. The residual map was constructed by subtracting a two-dimensional $\beta$-model fitted to the diffuse X-ray surface-brightness distribution of the cluster. In \citet{watson2026}, the residual image was smoothed with a Gaussian kernel of $\sigma = 7.5''$, while the original diffuse emission map was smoothed with $\sigma = 3''$. The residual highlights deviations from the smooth large-scale distribution of the ICM, revealing substructures associated with dynamical activity.
\renewcommand{\thefigure}{G.\arabic{figure}}
\setcounter{figure}{0} 
\begin{figure}[t!]
    \centering
    \includegraphics[width=\columnwidth]{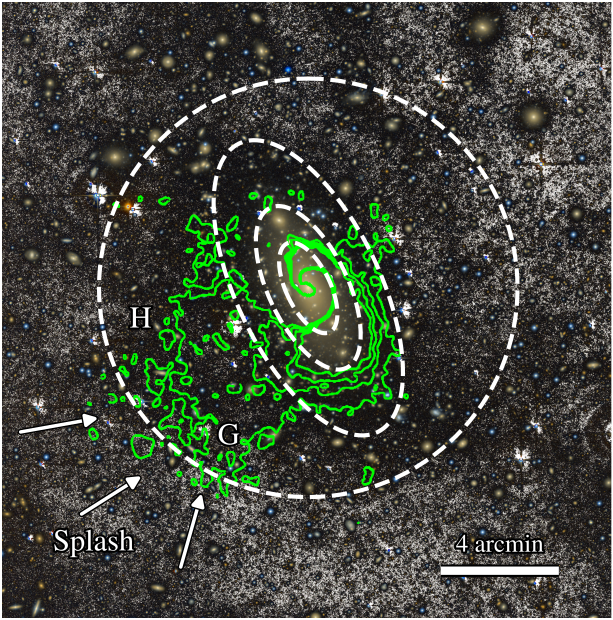}
    \caption{Ultra-deep optical image of IC~1101 with the residual X-ray emission contours from \citet{watson2026} overlaid in green. The residual corresponds to the 0.7--7 keV emission after subtraction of a best-fitting two-dimensional $\beta$-model and Gaussian smoothing with $\sigma=7.5''$. The white dashed ellipses and outer circle mark the four radial transitions identified in this work at $R=150$, $260$, $475$, and $620$~kpc, respectively. The LSB features labelled H and G are indicated, together with the ``splash'' feature reported by \citet{watson2026}.}
    \label{fig:xray_overlay}
\end{figure}

Remarkably, the X-ray feature identified by \citet{watson2026} as a ``splash'' exhibits a clear spatial coincidence with the LSB structures labelled H and G in Fig.~\ref{fig:full_color_image} and Fig.~\ref{fig:xray_overlay}. These optical features, detected after careful scattered light subtraction and background treatment, appear as extended diffuse components on the sides of IC~1101. Their projected alignment with the X-ray residual suggests that they may trace regions where the ICM has been recently perturbed, potentially by ongoing or past accretion events.

\citet{watson2026} further report that the spiral pattern is immediately evident in the residual X-ray emission, extending to $\sim600$~kpc from the cluster core. This scale is fully consistent with our independent photometric measurements of the outer stellar structures. In particular, we identify a radial feature at $R\sim620$~kpc, which coincides visually with the outer extent of the diffuse optical components H and G. This radius also matches the approximate termination of the X-ray spiral pattern. The agreement in spatial scale between the optical LSB envelope and the residual X-ray substructure strengthens the interpretation that both components trace the same large-scale dynamical processes operating in the cluster core.

This spatial correspondence supports the view that the outer stellar envelope and ICL around IC~1101 are dynamically connected to the recent assembly history of the cluster. The combined evidence from ultra-deep optical photometry and residual X-ray mapping points toward ongoing or relatively recent hierarchical accretion as a key driver of the observed structures.

\section{Conversion of Uson (1991) Luminosity to Stellar Mass}
\label{appendix:mass_conversion}

\renewcommand{\thefigure}{H.\arabic{figure}}
\setcounter{figure}{0} 
\begin{figure}[ht!]
    \centering
    \includegraphics[width =\columnwidth]{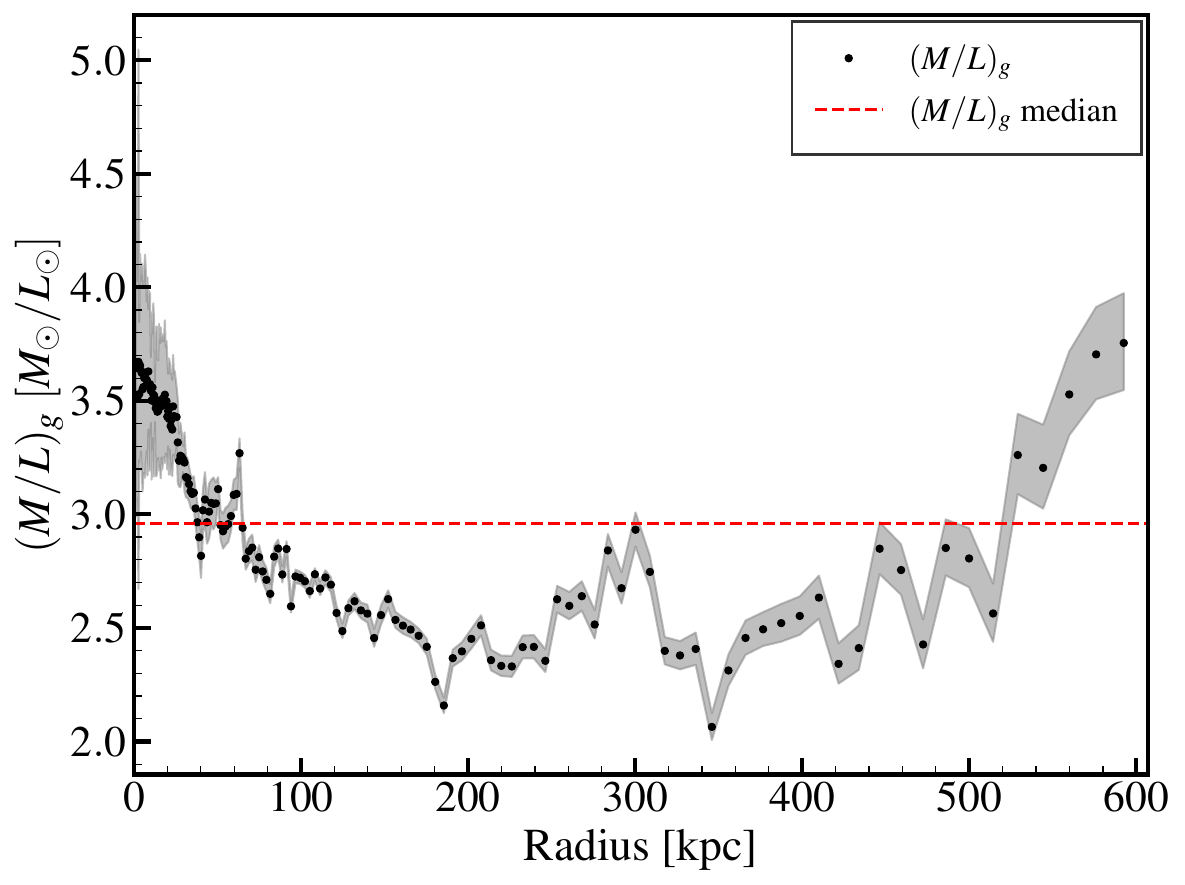}
    \caption{Radial profile of the $g$-band mass-to-light ratio $(M/L)_g$ derived from the colour profile of IC~1101, following the values of \citet{roediger2015}. The shaded region indicates the uncertainty at each radius. The red dashed line marks the median value within $R < 607$ kpc, which is used for the mass estimate based on the photometric luminosity from \citet{uson1991}.}    \label{fig:m_l_uson}
\end{figure}

To provide an independent comparison to our stellar mass estimate for IC~1101, we revisit the luminosity-based results reported by \citet{uson1991}. 
After rescaling their measurements to the cosmology adopted in this work ($h=0.7$), their integrated luminosity corresponds to $L \simeq 1.02 \times 10^{12}\,L_\odot$ within a radius of $d \simeq 607$~kpc. 
This radius corresponds to the effective radius defined by \citet{uson1991} as the geometric mean of two elliptical annuli, and the luminosity was obtained from a de Vaucouleurs profile fitted to the extended stellar halo of IC~1101.

The goal is to compare this luminosity-based estimate with our profile-based stellar mass measurement. 
We therefore adopt the same radius, $607$~kpc, and convert the luminosity into stellar mass using a mass-to-light ratio derived from our own photometric data.

To this end, we use the radial $(M/L)_g$ profile obtained from the colour-based prescription of \citet{roediger2015} (see Sect.~\ref{sec:radial_structure}), assuming a Chabrier \citet{chabrier2003} IMF for consistency with the rest of our stellar mass estimates. 
We compute the mean value of $(M/L)_g$ within $R < 607$~kpc and adopt it as a representative conversion factor. 
As shown in Fig.~\ref{fig:m_l_uson}, the average value in this range is $(M/L)_g = 2.9 \pm 0.3\,M_\odot/L_\odot$. 
Multiplying this value by the rescaled luminosity gives a stellar mass of $M_\star = 3.02 \times 10^{12}\,M_\odot$.

To estimate the uncertainty associated with this value, we follow a procedure analogous to that applied in our own profile-based mass derivation. 
We compute the standard deviation of the $(M/L)_g$ profile within the same radial range and use it to define an upper and lower bound for the total stellar mass. 
The resulting spread between these limits defines the error bar for the Uson-based estimate, yielding a final mass of $M_\star = (3.0 \pm 0.3) \times 10^{12}\,M_\odot$.
\end{appendix}

\end{document}